\def\figcond{1}     
\ifnum\figcond>0 
\documentstyle[preprint,aps,epsf]{revtex}
\else
\documentstyle[preprint,aps]{revtex}
\fi
\begin{document}
\preprint{DOE/ER/40427-02-N94}
\draft
\title{THE NUCLEON-NUCLEON POTENTIAL\\
IN THE CHROMO-DIELECTRIC SOLITON MODEL:  STATICS}
\author{W. Koepf, L. Wilets}
\address{
Department of Physics, FM-15, University of Washington,
Seattle, WA 98195, USA}
\author{S. Pepin and Fl. Stancu}
\address{
Universit\'e de Li\`ege, Institut de Physique B.5,
Sart Tilman, B-4000 Li\`ege 1, Belgium}
\date{\today}
\maketitle
\begin{abstract}

We study the nucleon-nucleon interaction in the framework of the
chromo-dielectric soliton model (CDM).  Here, the long-range parts of
the nonabelian gluon self-interactions are assumed to give rise to a
color-dielectric function which is parameterized in terms of an
effective scalar background field. The six-quark system is confined
in a deformed mean field through an effective non-linear interaction
between the quarks and the scalar field.  The CDM is covariant,
respects chiral invariance, leads to absolute color confinement and
is free of the spurious long range Van der Waals forces which trouble
non-relativistic investigations employing a confining potential.
Six-quark molecular-type configurations are generated as a function
of deformation and their energies are evaluated in a coupled channel
analysis. By using molecular states instead of cluster model
wave functions, all important six-quark configurations are properly
taken into account.  The corresponding Hamiltonian includes the
effective interaction between the quarks and the scalar background
field and quark-quark interactions generated through one gluon
exchange treated in Coulomb gauge.  When evaluating the gluonic
propagators, the inhomogeneity and deformation of the dielectric
medium are taken into account.  Results for the adiabatic
nucleon-nucleon potential are presented, and the various
contributions are discussed.  Finally, an outlook is given on how, in
the next stage of our investigation, the dynamical effects will be
incorporated by employing the generator coordinate method.
\end{abstract}

\bigskip
\newpage
\section{INTRODUCTION}

    The $N$-$N$ interaction is one of the most basic problems of
nuclear physics.  There exists extensive experimental information --
from $N$-$N$ scattering data and the properties of the deuteron --
but no single theoretical picture seems to be able to describe the
relevant physics for all internuclear distances.  In $N$-$N$
phenomenology, both relativistic and non-relativistic, one describes
the nucleons as elementary particles interacting through a two-body
potential which is either local or includes some nonlocality through
momentum-dependence in the interaction.  The general features of that
potential, i.e. the short-distance core and the long range
attraction, have been known for over forty years.

    Already in 1935, Yukawa \cite{Y35} suggested that the finite
range attraction was due to the exchange of an intermediate mass,
strongly-interacting particle, the subsequently discovered pion.
This led to the development of meson field-theoretic models which
today form the most accurate phenomenological description of the
$N$-$N$ interaction (see ref. \cite{M89} for an excellent overview).
In these models, one treats the nucleons as elementary particles with
an empirical form factor, and their interactions are usually carried
through one boson exchange (OBE) plus two pion exchange (TPE), where
the TPE is frequently simulated by a (fictitious) sigma meson.

    Within these descriptions, the long range ($r\agt$ 1.5 fm) part
of the $N$-$N$ interaction is controlled by one pion exchange, while
the intermediate range (0.5 fm $\alt r \alt$1.5 fm) attraction is
dominated by OBE and TPE.  The short range ($r\alt$ 0.5 fm) repulsion
is the ``mystery" region in such descriptions.  It has been described
by hard or soft cores, or form factors, both of the order of 0.5-0.8
fm, or by the exchange of vector mesons which have a range of
$1/m_\omega \ \approx\ $ 0.2 fm, a clear inconsistency.

    The advent of QCD and quark models has lifted the veil of mystery
from the short range $N$-$N$ problem exposing a new level of
simplicity.  However, the system is no longer just a two-body, but at
least a six-body, entity and more properly a field theoretical
problem.  The quark core of nucleons is of the order of 0.7 fm (an
rms radius of 0.5 fm) and one expects the quark substructure to be
effective within a range of $N$-$N$ separations of up to about 1 fm.

    A description of the nucleon-nucleon interaction within the
framework of quark degrees of freedom has been the subject of much
research.  The ideal venture would be a lattice gauge theory
calculation \cite{Latt}, but we are quite far from that stage,  and
therefore we have to rely on modeling.  We mention, non-exhaustively,
several different avenues which have been explored in that context:
non-relativistic constituent quark (potential) models \cite{L77},
relativistic current quark models as, for example, the MIT bag model
\cite{DT77} and various soliton models \cite{Schuh}, string models
and the topological Skyrme model \cite{W91,W92}. There are many
varieties under each category, and we will not attempt to review them
all here, but rather recommend the reader to the articles by Oka and
Yazaki \cite{O84}, Myhrer and Wroldsen \cite{M88} or Shimizu
\cite{S89}.

    In addition to the various Ans\"atze which have been employed to
model quark-quark interactions, one has to further distinguish
between static and dynamical calculations.  In the static
calculations, a local $N$-$N$ potential is obtained in
Born-Oppenheimer \cite{BO} (adiabatic) approximation from the energy
difference of a deformed six quark bag and two separated nucleons.
Non-adiabatic calculations yield a non-local $N$-$N$ interaction
through a consideration of the dynamics involved.  In the latter
category, usually the resonating group \cite{O81} or the generator
coordinate method \cite{C83} are applied.

    Quark models hold promise to give a good description for short
and intermediate range, but beyond, say, 1 fm the interaction,
although in principle describable in terms of quarks, is much more
easily represented by mesonic models, with the nucleonic substructure
giving rise to form factors for the meson-nucleon couplings.

    The ultimate object of our study is not only to reproduce the
two-body data, such as $N$-$N$ phase shifts and bound state
properties of the deuteron, but also to quantify the quark
substructure of nuclei.  With respect to the latter, we will describe
the collision process as an act of fusion followed by a separation
into three-quark clusters, and this will be used in conjunction with
the Independent Pair Model of nuclei to obtain quark structure
functions.  The main aspects of our current project can be described
as follows:

    1) We employ the chromo-dielectric soliton model \cite{F88,K88},
which respects covariance, contains absolute color confinement and is
free of the color Van der Waals problem \cite{VdW} (which is inherent
to most non-relativistic calculations).  In addition, one gluon
exchange is evaluated with a confined gluonic propagator.

    2) The six-quark wave function is expanded in terms of
``molecular" states \cite{St87}, including all configurations based
on the two lowest spatial single particle states.  This allows for
inclusion of basis states normally omitted in cluster model
calculations, and which have been demonstrated to be important in
decreasing the upper bound in variational calculations
\cite{St88,St89}.

    3) Dynamics will be handled through the generator coordinate
method \cite{GCM},  which leads to a set of coupled integral
equations.  It has been shown \cite{Schuh} that a significant part of
the short-range repulsion is due to dynamics, and the absence of a
repulsive core in some early calculations is now seen as an artifact
of the adiabatic approximation \cite{F83,H84}.  In addition, the
effective interaction is non-local in terms of the $N$-$N$ separation
parameter.

    4) In order to reproduce two-body properties, we will attach the
interaction we derive to a phenomenological local OBE potential
beyond a certain internuclear distance.  We could, however, also
consider extending our calculation more deeply into the OBE
intermediate range region by including quantum surface fluctuations
and introducing configurations of the form $q^7\bar q$ in addition to
our $q^6$ basis states.
    
    In this first of a series of papers, we are mostly concerned with
the introduction of the model and a presentation of the formalism we
use.  Therefore, we restrict ourselves to an adiabatic, or static,
approximation.  We calculate $<\alpha|H|\alpha>$, where $\alpha$ is
the separation or deformation parameter, including diagonalization
with respect to the various six-quark configurations.  We defer steps
(3) and (4) to subsequent papers in this series \cite{Fut}.

      The outline of this work is as follows.  In Sec. II we review
the chromo-dielectric soliton model, which in Sec. III is used to
generate single quark wave functions by means of a constrained mean
field calculation.  Sec. IV is devoted to the ``molecular" states
which form the basis for the six quark configurations we consider.
Sec. V describes the treatment of one gluon exchange and in Sec. VI
we present the results of our numerical calculations.  Finally, we
summarize, conclude and give an outlook on our future work in Sec.
VII.

\section{THE MODEL}

    The chromo-dielectric model \cite{F88,K88,Tang} is an evolution
of the Friedberg-Lee non-to\-po\-log\-i\-cal soliton model
\cite{F77}.  The Lagrangian is the same as the fundamental QCD
Lagrangian, supplemented by a scalar field $\sigma$ which is assumed
to simulate the gluonic condensate and other scalar structures which
inhabit the complicated physical vacuum.  It is assumed that the
scalar field, which has a non-vanishing vacuum expectation value
$\sigma_v$, parameterizes the bulk of the non-perturbative effects
which arise due to the non-linearity of QCD.  It furthermore governs
the chromo-dielectric properties of the medium.  The model Lagrangian
is covariant and, for massless quarks, satisfies chiral symmetry.  It
differs in that respect from most effective quark models -- such as
the MIT \cite{MIT}, Friedberg-Lee, or Nielsen-Patkos \cite{NP} models
-- which explicitly violate chiral symmetry through the interaction
of the quarks with some scalar field.  Although the model has its
basis in QCD, we regard it as phenomenological.

    The extra degrees of freedom introduced by the scalar field are
redundant.  In order to avoid double counting we do not include
diagrams which correspond to structures with the quantum numbers of
the $\sigma$ field.  Since the model parameters are readjusted at
each level of approximation to fit key physical data, one might hope
that as the level of sophistication of the calculations is increased,
one would find a decoupling of the $\sigma$ degrees of freedom and
would thus be left with pure QCD.  But, we are currently far from
that stage of sophistication.

    We treat the gluons in the Abelian approximation, which is
consistent with one gluon exchange.  The primary role of the $\sigma$
field is then to mediate the chromo-dielectric function,
$\kappa(\sigma)$, which, in turn, is designed to guarantee absolute
color confinement.  The scalar field, therefore, not only simulates
the non-linear and non-perturbative effects, but it also
parameterizes the color-dielectric properties of the space-time
continuum.

      The model Lagrangian density is given by
\begin{equation}
{\cal L}~~=~~{\cal L}_q~~+~~{\cal L}_\sigma~~+~~{\cal L}_G\ ,
\end{equation}
with
\begin{mathletters}
\begin{eqnarray}
{\cal L}_q~~&=&~~\overline{\psi}~\left(~i\gamma^\mu
D_\mu~-~m_q~\right)~\psi\ ,
\\ 
{\cal L}_\sigma~~&=&~~{1 \over 2}~\partial_\mu \sigma \partial^\mu
\sigma~- ~U(\sigma)\ ,
\\ 
{\cal L}_G~~&=&~~-~{1 \over 4}~\kappa(\sigma)~F_{\mu\nu}^c~F^{\mu\nu
c}\ .
\end{eqnarray}
\end{mathletters}

    Here, $m_q$ is the current quark mass matrix and, since there is
no direct coupling between the quarks and the $\sigma$-field, the
Lagrangian is chirally invariant for massless quarks.  For the rest
of this investigation we thus set $m_q \ \equiv \ 0$.  The gauge
field tensor is given by
\begin{equation}
F^a_{\mu\nu} =\partial_\mu A^a_\nu-\partial_\nu A^a_\mu+g_s
f^{abc}A^b_\mu A^c_\nu\ ,
\end{equation}
where the $f^{abc}$ are the SU(3) structure constants.  $U(\sigma)$
is the self-interaction energy of the scalar field and is taken to be
of the form
\begin{equation}
U(\sigma)~=~{a \over 2!}\sigma^2~+~{b \over 3!}\sigma^3~ +~{c \over
4!}\sigma^4~+~B\ .
\end{equation}
The quartic form of $U(\sigma)$ would assure renormalizability {\em if
$\kappa$ were a constant}.  Although this is convenient, it is not
demanded since we are already dealing with an effective theory.  The
bag pressure $B$ (which corresponds to the ``bag constant" of the MIT
model) is a function of the parameters $a,\ b$ and $c$ and is chosen
such that $U(\sigma)$ has a minimum and vanishes at the value
$\sigma=\sigma_v$, i.e. $U(\sigma_v)~=~U'(\sigma_v)~=~0\,$.  We
define
\begin{equation}
U''(\sigma_v)~\equiv~m_{GB}^2\ ,
\end{equation}
where $m_{GB}$ is identified with the mass of the lowest $0^{++}$
glueball state.  In order to guarantee absolute color confinement and
a regular behavior as $\sigma \rightarrow \sigma_v$, the dielectric
function must further satisfy
\begin{equation}
\kappa(\sigma_v)~=~\kappa'(\sigma_v)~=~\kappa'(0)~=~0
\qquad\hbox{and}\qquad 
\kappa(0)~=~1\ .
\end{equation}
We choose the form (with $x=\sigma/\sigma_v$)
\begin{equation}
\kappa(\sigma)=1\ + \theta(x)x^n(nx-(n+1))\ ,
\end{equation}
and set $n=2$ for our present investigation.

    Although the quarks are massless and there is no direct
quark-sigma coupling, they still acquire a self-energy -- and hence
an effective mass  -- through their interactions with the gluon
field.  Furthermore, the  gluonic propagator depends on $\sigma$
through $\kappa(\sigma)$ and is thus also ``confined".

    This confinement mechanism has been studied for two particular
cases: (1) a uniform dielectric function \cite{K88} and (2) a cavity
model in which $\kappa$ is unity at the center and goes to zero
outside the bag \cite{Tang}.

      In the preliminary discussion \cite{F88} employing a uniform
dielectric, confinement was exhibited explicitly from a calculation
of the quark self-energy in the limit of very heavy, i.e. fixed,
quarks.  In calculations with zero mass quarks \cite{K88}, it was
shown that the quarks acquire a ``dynamic" mass below some critical
$\kappa_c$.  The corresponding self-energy displays an asymptotic
form which increases as $\kappa$ decreases and becomes infinite as
$\kappa \rightarrow 0$.  This corresponds to a realization of {\em
spatial confinement}, since this mass is ``color blind." It was shown
that a massless Goldstone pion arises as a direct consequence of
spatial confinement, and that the successful cloudy bag model emerges
quite naturally in this formalism.  The resulting quark propagator
exhibits momentum dependence and hence non-locality.

      {\em Color confinement}, on the other hand, arises through the
enclosure of the quark cavity by the physical vacuum where the
dielectric function goes to zero.  The gluon field energy is thus
infinite if the quark structure in the cavity is not in a color
singlet state. Note that $\kappa\rightarrow 0$ also ensures that
there are no spurious color Van der Waals forces \cite{VdW} since
then the stress-energy tensor vanishes in the vacuum.  In a study
\cite{Tang} employing a cavity model, the non-local quark propagator
was calculated by solving the Schwinger-Dyson equation.

    We utilize the results of these studies and introduce an
effective quark mass which is designed to simulate spatial
confinement.  We thus add an {\em effective coupling} between the
quarks and the scalar field.  This destroys chiral invariance (which,
in turn, is restored by the emergence of the Goldstone pion).  The
form of the added term is
\begin{equation}
{\cal L}_{q\sigma}~~=~~-~g_{eff}(\sigma)~\overline{\psi}~\psi\ ,
\end{equation}
with
\begin{equation}
g_{eff}(\sigma)~=~g_0~\sigma_v \left(~{1 \over
\kappa(\sigma)}~-~1~\right)\
\end{equation}
motivated by comparison with the results of ref. \cite{K88}.  It
turns out, however, that the exact form of that coupling term does
not seem to be too important.

    Color confinement is realized by incorporating one gluon exchange
(OGE) matrix elements both for the quark self-energy and for the
mutual interaction terms.  Since we work in the Coulomb (or
transverse) gauge, where $\mbox{\boldmath $\nabla$}
\cdot (\kappa {\bf A}^c)=0$, the time component $A_0^c$, which 
is responsible for color confinement, is instantaneous and
frequency-independent.  The effective Hamiltonian can be written as
\begin{equation}
H = \int\!d^3\!{\bf r}\, {\cal{H}}({\bf r})\ ,
\end{equation}
where
\begin{eqnarray}
{\cal{H}}({\bf r}) =\psi^\dagger\!({\bf r})
\Big\{\,\mbox{\boldmath $\alpha$}
\cdot[{\bf p} - 
{\scriptstyle{1\over 4}} g_s \, \mbox{\boldmath $\lambda$}^c \, {\bf
A}^c({\bf r})] \, + \,
\beta \, g_{eff}\bigl(\sigma({\bf r})\bigr) \, &+& \, 
{\scriptstyle{1\over 4}} g_s \, \mbox{\boldmath $\lambda$}^c \,
A_0^c({\bf r})
\Big\}\psi({\bf r}) \\ 
&+& \, {\scriptstyle{1\over 2}} \!
\left(\dot\sigma({\bf r})\right)^2 \, + \,
{\scriptstyle{1\over 2}} |\mbox{\boldmath $\nabla$}\sigma ({\bf
r})|^2 \, + \, U\!(\sigma)\ , \nonumber
\end{eqnarray}
and where the $\mbox{\boldmath $\lambda$}^c$ are Gell-Mann's color
$SU(3)$ matrices normalized so that $\sum_c\mbox{\boldmath
$\lambda$}^c\cdot
\mbox{\boldmath $\lambda$}^c=16/3$.  This must be
supplemented by the field equations for $A_0^c$ and ${\bf A}^c$,
to be given in Sec.  V.

    In order to fit the five parameters of the chromo-dielectric
soliton model, $a$, $b$ and $c$ in $U(\sigma)$, $g_0$ in $g_{eff}$
and the strong coupling constant $\alpha_s=g_s^2/4\pi$, we construct
self-consistent solutions for the nucleon.  Employing the coherent
state approximation, we treat the scalar field classically.  In
addition, gluonic terms are dropped when determining the quark wave
function, $\psi({\bf r})$, or the scalar field, $\sigma({\bf r})$.
Making the Ansatz (see ref. \cite{H85} for more details) that each
quark is in an $s$-state, we have
\begin{equation}
\psi_m({\bf r})~~=~~\left( \begin{array}{c} 
u(r) \\ i~\mbox{\boldmath $\sigma$}
\cdot {\bf \hat r}~v(r) \end{array} \right)~\chi_m
\end{equation}
for the quark wave function, where $\chi_m$ is a Pauli spinor.  This
gives the mean field equations of motion
\begin{mathletters}
\begin{eqnarray}
{du \over dr}~~&=&~~-~(g_{eff}(\sigma)+\epsilon)~v\ ,
\\         
{dv \over dr}~~&=&~~-~{2v \over r}~-~(g_{eff}(\sigma)-\epsilon)~u\ ,
\\         
{d^2\sigma \over dr^2}~+~{2 \over r}{d\sigma \over dr}~~&=&~~
{dU(\sigma) \over d\sigma}~+~ n_q{dg_{eff}(\sigma) \over
d\sigma}(u^2-v^2)\ ,
\end{eqnarray}
\end{mathletters}
which are solved self-consistently, with $n_q=3$ for the nucleon.
This yields the solitonic quark wave function and the corresponding
scalar field, from which then the quantities of physical interest are
calculated while incorporating certain approximate recoil corrections
\cite{recoil}; the latter can be compared with methods using
projection and boost \cite{boost}.  Of the five parameters involved
-- $a$, $b$, $c$, $g_0$ and $\alpha_s$ -- three are fixed by fitting
the nucleon mass, the $\Delta$ mass and the proton rms charge radius
(0.83~fm). This leaves two free parameters, for which we choose the
dimensionless quantities $f=b^2/ac$ and $c$.

    In Fig. 1 we show the self-consistently determined scalar field,
$\sigma/\sigma_v$, and the corresponding self-interaction potential,
$U(\sigma)$, for three different parameter sets.  The solid line
corresponds to $f=3$, for which the bag pressure $B$ vanishes and
which creates hard bags with a thin surface. The dashed line
corresponds to $f=\infty$, for which the quadratic term in
$U(\sigma)$ disappears and $\sigma=0$ turns from a second minimum to
an inflection point and which generates soft bags with a thick
surface.  The dot-dashed line corresponds to $f=3.2$.  For all sets,
$c$ is kept at a value of $10000$, which yields a reasonable figure
for the glueball mass (1.5 GeV $\alt m_{GB} \alt$ 2.0 GeV).  In
general, for increasing $f$ or $c$, the bag gets softer, the glueball
mass and the bag pressure increase, the agreement in the axial vector
coupling $g_A$ -- which is inherently too small by about 10 percent
-- improves, but the proton's magnetic moment $\mu_p$ -- which is
also consistently underestimated -- grows to differ more from its
experimental value.  Better agreement with experiment can be achieved
by using projection and boost \cite{boost}.  Table I gives an
overview over the corresponding quantities for the various parameter
sets under consideration.

    As already remarked, we work in the one gluon exchange
approximation.  Since for both the nucleon and the $\Delta$, all of
the quarks are in the same spatial state and the entire system is a
color singlet, the total -- mutual plus self -- color-electrostatic
energy is zero.  The color-magnetic interaction, on the other hand,
is responsible for the $N$-$\Delta$ mass splitting.  In general, part
of this energy difference should be attributed to the different pion
dressing of the nucleon and the $\Delta$, but since at our present
level of approximation the soliton does not contain any pionic
effects, we disregard this contribution.  As usual \cite{B851}, the
magnetic self-energy contribution is neglected here and no
intermediate excitations of the quarks into higher spatial orbitals
are taken into account.  The evaluation of the $N$-$\Delta$ mass
splitting allows the adjustment of the strong coupling constant,
$\alpha_s$. Hereby, a ``confined" gluonic propagator is used, i.e.
the explicit dependence of the gluonic field equations on the
dielectric function $\kappa(\sigma)$ is taken into account.  For
details see Sec. V and ref. \cite{B851}.  The results are given in
table I.  For comparison, we also show the values of $\alpha_s$ which
we obtain by using a free propagator, i.e.  by setting $\kappa\equiv
1$.

    Another quantity which depends on the gluonic interactions is the
string tension $\theta$.  Phenomenologically, it is obtained from
fits to heavy quarkonium spectra as the coefficient of the linear
term in non-relativistic $q\bar q$ potentials ($\theta_{exp} \approx$
913 MeV \cite{C79}).  In the soliton bag model, it can be calculated
by considering a flux tube, which is a cylindrical structure of
scalar and gluon fields generated by a quark and an antiquark pair of
color charges at infinitely large separation.  In particular, by
minimizing the energy of such a system, we obtain the spatial form of
the scalar and color-electric fields in the tube, and the string
tension.  Alternatively, we can adjust the parameters of the model to
obtain the ``experimental" string tension.  Results for $\alpha_s$
obtained in this way are also given in Table I.  For further details
we refer the reader to ref. \cite{B851}.

\section{CONSTRAINED MEAN FIELD APPROXIMATION}

    The starting point of any evaluation of the multi-dimensional
potential energy surface and the input to any calculation employing
the generator coordinate method \cite{GCM} is a wave function which
is characterized by a set of deformation parameters, which in the
following will be denoted collectively as $\mbox{\boldmath
$\alpha$}$.  In our case, the $\mbox{\boldmath $\alpha$}$ describe
the static configuration of a system of six quarks and the
corresponding deformed scalar $\sigma$-field, which is treated
quantum mechanically through the coherent state approach.
Consideration of various possible six-quark configurations (see Sec.
IV for more details) allows for each deformation the construction of
a complete basis, which is indicated by a set of state vectors
$|\mbox{\boldmath $\alpha$},n\!>$.

    In general, these state vectors are generated by means of a
constrained mean field calculation, i.e. by  extremizing the
expectation value of the total Hamiltonian, as given by
$<\!\mbox{\boldmath $\alpha$}|H|\mbox{\boldmath $\alpha$}\!>$, with
respect to a variational wave function for the quarks and a coherent
state for the scalar field,  subject to the constraints
\begin{equation}
<\!\mbox{\boldmath $\alpha$}|
\mbox{\boldmath $Q$}|\mbox{\boldmath $\alpha$}\!> =
\mbox{\boldmath $Q$}_0 \ ,
\end{equation}
where the $\mbox{\boldmath $Q$}$ are some moments of the quark
distribution as defined through
\begin{equation}
\mbox{\boldmath $Q$} 
= \int\!\overline\psi({\bf r})\,
\mbox{\boldmath $q$}({\bf r})\,\psi({\bf r})\,d^3\!{\bf r}\ ,
\end{equation}
for some chosen set $\mbox{\boldmath $q$}({\bf r})$.  In the above,
we have dropped the label ``$n$" for simplicity, and we here limit
ourselves to a one-dimensional parameter space, i.e. we consider
zero-impact trajectories (or central collisions) only.  The
constrained mean field equations then assume the form
\begin{eqnarray}
\Big\{ \mbox{\boldmath $\alpha$}
\cdot{\bf p}+\beta\big[g_{eff}(\sigma({\bf r}))
-\lambda\,q({\bf r})\big]-\epsilon_\mu\Big\} \psi_\mu~&=&~0\ ,
\\       
-\nabla^2 \sigma+{dU(\sigma) \over d\sigma}+{dg_{eff}(\sigma)
\over d\sigma} <\overline\psi \psi> ~&=&~0\ .           
\end{eqnarray}
where $<\overline\psi \psi>$ is the six-quark scalar density and
$\lambda$ is a Lagrange multiplier imposing the subsidiary condition.
All gluonic terms have been dropped from the above equations, and the
label $\mu$ identifies the different single-particle quark states.

    Instead of specifying the constraint function $q({\bf r})$
explicitly and solving the above pair of equations simultaneously and
self-consistently, it is easier and actually more physical to specify
the function
\begin{equation}
\big[g_{eff}(\sigma({\bf r}))-\lambda\, q({\bf r})\big]~\equiv~
{\cal{V}}_\alpha({\bf r})~\equiv~g_{eff}(\sigma_\alpha({\bf r}))
\end{equation}
for each value of the collective deformation parameters $\alpha$.
${\cal{V}}_\alpha({\bf r})$ plays the role of an external potential
generating the wave function for the quarks.  It is expressed in
terms of a function of some scalar field with a prescribed
deformation, $\sigma_\alpha({\bf r})$.  In employing this scalar
field with a particular deformation, we explicitly give up
self-consistency between the quark wave functions and the scalar
field.  However, this lack of self-consistency is necessarily
inherent to any constrained calculation.

    We construct the field $\sigma_\alpha({\bf r})$ by folding a
Yukawa shaped smoothing function with the union (for $\alpha > 0$) or
the intersection (for $\alpha < 0$) of two spheres \cite{Schuh},
whose centers are separated by a distance $|\alpha|$, i.e.
\begin{eqnarray}
\sigma_\alpha({\bf r})~&=&~\sigma_v~-~
\sigma_0 \int\!T_\alpha({\bf r^\prime}) 
f(|{\bf r}-{\bf r^\prime}|)\,d^3\!{\bf r^\prime}\ ,
\\         
T_\alpha({\bf r})~&=&~\left\{ \begin{array}{ll}
\theta (R(\alpha) - |{\bf r}-\hat{\bf z}\alpha/ 2 |) &
\mbox{for $z \geq 0$}\ , \\
\theta (R(\alpha) - |{\bf r}+\hat{\bf z}\alpha/ 2 |) &
\mbox{for $z  <   0$}\ , \end{array} \right.
\\         
f(r)~&=&~{\Gamma^2 \over 4\pi}{e^{-\Gamma r} \over r}\ ,
\end{eqnarray}
where $\sigma_v$ is the vacuum expectation value of the scalar field
and where $\alpha>0$ corresponds to prolate deformations and
$\alpha<0$ to oblate deformations.

    The geometrical parameters, $R$, $\Gamma$ and $\sigma_0$, are
determined from the scalar field $\sigma_N({\bf r})$ of a
self-consistent solution for the nucleon, as discussed in the last
section, such that the corresponding solitonic scalar field of two
free nucleons is well approximated at asymptotic deformations, i.e.
\begin{equation}
\sigma_{\alpha\rightarrow\infty}({\bf r})~\rightarrow~
\sigma_N({\bf r}-\hat{\bf z}\alpha/ 2)~+~
\sigma_N({\bf r}+\hat{\bf z}\alpha/ 2)-\sigma_v\ .
\end{equation}
In order to select a definite path in configuration space, as a first 
approximation the field strength $\sigma_0$ and the surface parameter
$\Gamma$ are kept constant and the radius $R=R(\alpha)$ is varied in
such a way that the volume of the scalar field of the six-quark
cavity is independent of the deformation parameter and remains fixed
at the value of two nucleonic volumes \cite{Schuh}.  In Fig. 2, we
show the field $\sigma_\alpha({\bf r})$ obtained in that manner for
four different $\alpha$'s.

    Selecting this particular path in the geometrical configuration
space spanned by $R$, $\Gamma$ and $\sigma_0$ is equivalent to
treating the scalar $\sigma$-field as an incompressible liquid.  In
order to check this approximation, we constructed self-consistent
stationary eigen-states of the total Hamiltonian for a spherically
symmetric scalar field, which corresponds to united bags, i.e. to
$\alpha=0$.  Our findings, which will be discussed in Sec. VI, show
that the quality of the ``constant volume" approximation is quite
remarkable.

    The potential ${\cal{V}}_\alpha({\bf r})$ serves to generate a
set of single-particle quark states, $\psi^\alpha_n({\bf r})$, which
are determined from the eigenvalue equation (16).  Here, we limit
ourselves to the lowest states of positive and negative parity,
denoted by $|\sigma\!>$ and $|\pi\!>$ respectively, and to values for
the single-particle magnetic quantum numbers of $m=\pm 1/2$.  The
corresponding eigenenergies, $\epsilon_\sigma$ and $\epsilon_\pi$,
are shown in Fig. 3 for two particular parameter sets ($c=10000$;
$f=3$ and $f=\infty$) adjusted to the standard properties of the
nucleon.  This figure depicts the increasing binding of the positive
parity state, $|\sigma\!>$, for small $\alpha$ as well as the
convergence of both levels for separating bags. As $\alpha\to\infty$,
the two states become linear combinations of $R$ and $L$,
corresponding to an $s$-state in each bag.  As $\alpha \to 0$, the
single-particle states evolve to $\sigma \to s_{1/2}$ and $\pi \to
p_{3/2},m=\pm 1/2$.

\section{QUARK MOLECULAR BASIS STATES}

    The classification and construction of antisymmetric six-quark
basis states is a central part of any study of the $N$-$N$ system in
terms of quark degrees of freedom.  In principle, the choice of a
particular basis is irrelevant, as long as a sufficiently large
number of configurations is included.  However, in practice we have
to content ourselves with a relatively small subset of states, and
therefore need to make sure that these include the configurations
which supposedly dominate the exact ground state of the corresponding
Hamiltonian.

    For the construction of the six quark states, incorporating all
possible degrees of freedom, namely color (C), orbital motion (O),
spin (S) and isospin (T), we use a classification scheme based on
$SU(4)$ spin-isospin symmetry, as introduced by Harvey \cite{H81}.
Here, one first combines the color singlet $SU(3)$ function, as
denoted by the partition [222]$_C$, with an orbital function of a
specific symmetry, say $[f]_O$. This leads to a state of combined
symmetry $[\tilde f']_{OC}$.  The latter is then coupled to a
spin-isospin $SU(4)$ function with the {\em dual} symmetry,
$[f']_{TS}$.  This yields a configuration which is totally
antisymmetric with respect to the interchange of any pair of
particles, i.e. the has Young symmetry $[1^6]$.  We thus arrive at
the notation
\begin{equation}
\psi_6~=~\left(~\left([f]_O\;[222]_C\right)_{[\tilde f']_{OC}}\;
[f']_{TS}~ \right)_{[1^6]}\ .
\end{equation}

    Using ``fractional parentage coefficients," wave functions like
(23) can be written as linear combinations of products of
antisymmetric states of the first $n-1$ particles and the last
particle or, alternatively, of the first $n-2$ particles and the last
pair \cite{H81}. This helps to reduce the six-body matrix elements of
the one- and two-body Hamiltonian to linear combinations of one- and
two-body matrix elements, by taking the orthogonality of the $n-1$
and $n-2$ particle states of distinct symmetries into account.

      The only sectors which are compatible with $L=0$ $N$-$N$
partial waves are T=0, S=1 and T=1, S=0 \cite{Ring}.  The relevant
orbital symmetries are then $[f]_O=[6]$ and $[42]$ if one assumes
that each nucleon is asymptotically in an orbital $[3]$ state.   For
the spin-isospin function, on the other hand, quite a few $SU(4)$
representations, $[f']_{TS}$, would in principle be available, and
they can, for example, be found in Harvey's work \cite{H81}.  Yet
among them only the $[f']_{TS}=[6]$, $[51]$, $[42]$ and $[33]$  --
which are labeled with ``asterisks" in Harvey's article -- evolve
into asymptotic di-baryon states for large inter-nucleon separations.
As all other $[f']_{TS}$ states -- the ``non-asterisked" ones --
couple very weakly to the latter \cite{St88,St89}, they can safely be
disregarded for the $N$-$N$ problem.

    The novelty with respect to Harvey's scheme -- and similar other
studies -- lays in the choice of the orbital share of the wave
function.  In most previous calculations, the ``cluster model" has
been used -- see ref. \cite{M88} for a review -- which describes the
orbital degrees of freedom in terms of two separate three-quark
clusters centered at the locations of the two respective nucleons,
denoted in the following as $|R\!>$ and $|L\!>$.  In this
investigation, on the other hand, we use ``molecular orbitals"
\cite{St87} where the spatial single-particle states are wave
functions of a static single-particle Hamiltonian, such as obtained
from constrained Hartree-Fock or soliton mean field theories.  In our
case, the respective single-quark basis states are the two lowest
orbitals of either parity, $|\sigma\!>$ and $|\pi\!>$, discussed in
the last section.  It is obvious that the molecular states are
orthogonal at any separation $\alpha$, whereas the cluster model
states are not. The latter even overlap completely for vanishing
inter-nucleon separation, $<\!R|L\!>\!|_{\alpha=0}=1$.  Also, the
molecular states are natural to a mean field description as, for
example, the one employed in this work.  Note also that in the
cluster model, the limit $\alpha \to 0$ requires special care in the
normalization of the various symmetry configurations \cite{St89}.
Otherwise some contributions are mistakenly left out, as was the case
e.g. in refs. \cite{F83} and \cite{H81}.

    For large separations between the nucleonic bags, $|\sigma\!>$
and $|\pi\!>$ become degenerate and turn into orthogonal combinations
representing the lowest s-state located in either bag.  For the study
of the $N$-$N$ scattering problem, it is convenient to construct
orthogonal, pseudo-right and pseudo-left orbitals, $|r\!>$ and
$|\ell\!>$, respectively as
\begin{equation} \left. \begin{array}{c}
|r\!> \\ |\ell\!> \end{array} \right\}~=~ {|\sigma\!>~\pm~|\pi\!>
\over \sqrt{2}}
\ . 
\end{equation}
The $|r,\ell\!>$ molecular states recover the $|R,L\!>$ cluster model
states at large separation, but at finite deformation their behavior
is obviously very different from the cluster model states.  This
proves to be important for the short range part of the $N$-$N$
interaction \cite{St88,St89}.

    The transformation from the $|r,\ell\!>$ to the $|\sigma,\pi\!>$
scheme for the relevant six quark basis states is given in table 1 of
ref. \cite{St87}.  This table also suggests that, due to the complex
structure of those configurations, in all practical calculations it
is much simpler to work with the $|r,\ell\!>$ basis states instead.
One can then eventually return to the $|\sigma,\pi\!>$ configurations
at the level of one- or two-body matrix elements.  In refs.
\cite{St88} and \cite{St89}, results for the united six-quark bag
obtained with cluster model wave functions were compared with
corresponding calculations employing a molecular basis and, in
particular, the constituent quark model and the MIT bag model were
investigated in that context.  In both cases, the authors found that
at zero separation the ground state energies were substantially
lowered through the use of the molecular orbitals.  The reason for
this is that configurations of the type $|\sigma^n\pi^{6-n}(n\not=
3)\!>$, which are missing in a cluster model basis, proved to be
quite important.  For the relativistic current quark model
\cite{St89}, the most relevant states for the isospin-spin channels
$(TS)=(01)$ or $(10)$ turned out to be:
\begin{eqnarray}
|1\!>~&=&~|NN\!>\ ,                          \nonumber \\
|2\!>~&=&~|\Delta\Delta\!>\ ,                \nonumber \\
|3\!>~&=&~|CC\!>\ ,                          \nonumber \\
|4\!>~&=&~|42^+ [6]_O~ \{33\}_{TS} \!>\ ,    \\ 
|5\!>~&=&~|42^+ [42]_O \{33\}_{TS} \!>\ ,    \nonumber \\ 
|6\!>~&=&~|42^+ [42]_O \{51\}_{TS} \!>\ ,    \nonumber \\ 
|7\!>~&=&~|51^+ [6]_O~ \{33\}_{TS} \!>\ ,    \nonumber
\end{eqnarray}
where the notation of ref. \cite{St87} was used.  The first three
form the ``physical" basis in Harvey's investigation of the $N$-$N$
interaction \cite{H81}, and they contain solely configurations of the
type $|r^3\ell^3\!>$.  The other four are of the form
$|r^4\ell^2+r^2\ell^4\!>$ or $|r^5\ell+r\ell^5\!>$, denoted as $42^+$
and $51^+$, respectively.  Configurations of that type (with $R$ and
$L$ replacing $r$ and $\ell$) do not occur in a cluster model basis.
In accordance with the findings of ref. \cite{St89}, we choose the
seven states of Eq. (25) to be the basis of our truncated Hilbert
space in which the Hamiltonian, as described in Sec. III (one-body
part) and Sec. V (two-body part), has been diagonalized.

\section{ONE GLUON EXCHANGE}

    We treat quark-gluonic interactions in the one gluon exchange
approximation.  At this level, we do not encounter the problem of
double-counting, since colorless structures which are already
represented by the scalar field begin with two-gluon exchange or the
excitation of $q\overline q$ pairs.  Higher order effects which arise
due to the non-Abelian character of QCD are also assumed to be
simulated by the scalar field.

    In addition, the non-Abelian terms in the QCD gauge field tensor,
$F^a_{\mu\nu}$ of Eq. (3), have been neglected.  The field equations
therefore linearize and become identical to Maxwell's equations in an
inhomogeneous medium, with the exception, however, that now all the
field operators and currents contain Gell-Mann's color $SU(3)$
matrices.  In this approximation we find from Eqs. (2a), (2c) and (3)
\begin{equation}
\partial^\mu \left( \kappa(\sigma) \left( \partial_\mu A^c_\nu - 
\partial_\nu A^c_\mu \right) \right)~=~J_\nu^c\ , 
\end{equation}
where the total quark color-current operator is
\begin{equation}
J_\nu^c~=~ {g_s \over 2} \, \overline \psi \gamma_\nu \mbox{\boldmath
$\lambda$}^c \psi
\ ,
\end{equation}
with $g_s=\sqrt{4\pi\alpha_s}$.  The gluonic fields are explicitly
affected by the scalar field through $\kappa(\sigma)$.  As the
dielectric is constructed in such a way as to ensure absolute color
confinement -- i.e. it vanishes in the vacuum -- the resulting
gluonic propagators will also be ``confined" and there will be no
gluons propagating outside the solitonic bags.

    In order to solve the field equations, we choose the Coulomb, or
transverse, gauge,
\begin{equation}
\mbox{\boldmath $\nabla$}\cdot \left(\kappa {\bf A}\right)~=~0\ ,
\end{equation}
which decouples $A_0$ in (26) through
\begin{equation}
-~\mbox{\boldmath $\nabla$} \cdot \kappa~\mbox{\boldmath $\nabla$}
A_0~=~J_0\ .
\end{equation}
The field equation for the space components of $A_\mu$ reads
\begin{equation}
\kappa \, \partial_t^2 {\bf A} ~-~ \nabla^2 \kappa {\bf A} ~+~ 
\mbox{\boldmath $\nabla$} \times \left( {\bf A} \times 
\mbox{\boldmath $\nabla$}\kappa \right) ~=~ 
{\bf J}_t \ ,
\end{equation}
where the transverse current, which satisfies $\mbox{\boldmath
$\nabla$}\cdot{\bf J}_t=0$, is defined by means of
\begin{equation}
{\bf J}_t ~=~ {\bf J} ~-~ \kappa \, \nabla \, \partial_t A_0 \ .
\end{equation}

    We note that due to the scalar nature of the medium, the field
equations are diagonal in the color indices, which have hence been
omitted in Eqs. (28) to (31).  From these equations we can
furthermore deduce the mutual and self-interaction energies between
the quarks which arise due to the OGE through
\begin{equation}
H_{OGE}~=~{1\over 2} \int\!d^3\!{\bf r} \, J_\mu^c \, A^{\mu c} \ ,
\end{equation}
and finally evaluate their contributions to the one-body
(self-interactions) and two-body (mutual interactions) parts of the
effective Hamiltonian.

    The corresponding matrix elements are evaluated by first
determining the ``gluon propagator in medium". We follow here
Bickeboeller et al. \cite{B85}, as corrected subsequently by Tang and
Wilets \cite{Tang2}, where, however, the corrections in ref.
\cite{Tang2} do not affect the matrix elements needed here.  In those
papers, the respective Green's functions of the differential
equations (29) and (30) were calculated by making an expansion in
either spherical harmonics (for the scalar Green's function $G({\bf
r},{\bf r^\prime})$) or vector spherical harmonics  (for the tensor
Green's function $G^{ij}({\bf r},{\bf r^\prime},\omega)$).  The
resulting coupled differential equations were then solved numerically
in an angular momentum representation.  Since Eq. (29) for the time
component $A_0$ contains no explicit time derivatives, the
corresponding Green's function, $G({\bf r},{\bf r^\prime})$, is
frequency independent and the field $A_0$ is consequently
instantaneous.  For further details the reader may consult refs.
\cite{B85} and \cite{Tang2}.

    The part of the OGE interactions that arises from the time
component of the gluonic field $A_0^c$ and which is mediated through
the scalar Green's function is responsible for the realization of
color confinement, and the part of the OGE interactions that stems
from the spatial components of the gluonic field ${\bf A}^c$ as cast
by the tensor Green's function $G^{ij}({\bf r},{\bf r^\prime},
\omega)$ generates the color-magnetic hyperfine interaction which, in
turn, produces the $N$-$\Delta$ mass splitting.  As usual, the
self-interaction terms have been included in the time part of the OGE
and have been neglected for the spatial contributions.  This is in
accordance with the minimal self-energy prescription of the MIT bag
model \cite{W84}. Only when taking the color-electrostatic
self-energy diagrams arising from the time component of the gluonic
field into account, the color-electrostatic interaction between two
well separated nucleonic color singlets vanishes, as is required by
color neutrality \cite{DT77}.

    In addition, the OGE matrix elements which arise from the tensor
Green's function, i.e. from the color-magnetic hyperfine interaction,
are somewhat smaller than the ones which are generated by the
color-electrostatic interaction. The reason for this reduction is
that the latter always involve the ``small" lower components of the
relativistic quark spinors.  Inasmuch as the magnetic interaction is
not directly involved in color confinement, we simplify our
calculations by using a free ($\kappa=1$) tensor propagator,
$G_f^{ij}({\bf r},{\bf r^\prime}, \omega)$, and an effective
$\alpha_s \to \alpha_s^{free}$ adjusted to yield the experimental
$N$-$\Delta$ splitting.

    To leading order in the strong coupling constant, the gluonic
field energy contribution to the effective Hamiltonian can be written
as a sum of terms each involving a $c$-number,
configuration-independent energy and an operator which depends on
color, spin, flavor and orbital quantum numbers \cite{DT77}.  We find
\begin{eqnarray}
< H_{OGE} >~=~& &V_0^{pqrs} < b_p^+ \mbox{\boldmath $\lambda$}^c b_q
\, b_r^+ \mbox{\boldmath $\lambda$}^c b_s > ~+~ V_z^{pqrs} < b_p^+
\sigma_z \mbox{\boldmath $\lambda$}^c b_q \, b_r^+
\sigma_z \mbox{\boldmath $\lambda$}^c b_s > 
\nonumber \\ 
& &+~V_\bot^{pqrs} < b_p^+ \mbox{\boldmath $\sigma$}_\bot
\mbox{\boldmath $\lambda$}^c b_q \cdot b_r^+ 
\mbox{\boldmath $\sigma$}_\bot \mbox{\boldmath $\lambda$}^c b_s > ~+~ 
\sum_q V_0^{pqqp} < b_p^+ \mbox{\boldmath $\lambda$}^c \, 
\mbox{\boldmath $\lambda$}^c b_p > \\         
& &+~\sum_q V_z^{pqqp} < b_p^+ \sigma_z \mbox{\boldmath $\lambda$}^c
\,
\sigma_z \mbox{\boldmath $\lambda$}^c b_p > 
~+~ \sum_q V_\bot^{pqqp} < b_p^+ \mbox{\boldmath $\sigma$}_\bot
\mbox{\boldmath $\lambda$}^c \cdot 
\mbox{\boldmath $\sigma$}_\bot \mbox{\boldmath $\lambda$}^c b_p > 
\ , \nonumber
\end{eqnarray}
where the first three terms on the right hand side correspond to the
two-body mutual interactions, depicted in Fig. 4a, and the last three
correspond to the one-body self-interactions, as shown in Fig. 4b.
The $b$'s ($b^+$'s) are annihilation (creation) operators for the
single-quark states $|\sigma\!>$ and $|\pi\!>$ which were introduced
in Sec. III.  The configuration-dependent matrix elements, $< b^+
\ldots b >$, are evaluated with the method of ``fractional parentage
coefficients", which was outlined in the last section.  They generate
an explicit mixing between the various six-quark configurations under
consideration.
  
    The $V^{pqrs}$, on the other hand, are configuration independent
interaction energies,
\begin{eqnarray}
V_0^{pqrs}~&=&~\sum_{m,m'} {\cal V}^{\, pm;qm}_{\, rm';sm'} \ ,
\nonumber \\ V_z^{pqrs}~&=&~\sum_{m,m'} (-)^{m-m'} {\cal V}^{\,
pm;qm}_{\, rm';sm'} \ , \\ V_\bot^{pqrs}~&=&~\sum_{m \ne m'} {\cal
V}^{\, pm;qm'}_{\, rm';sm} \ ,
\nonumber
\end{eqnarray}
where each individual summand
\begin{eqnarray}
{\cal V}^{\, p m_p;q m_q}_{\, r m_r;s m_s}~=~{g_s^2 \over 32}
\int\!d^3\!{\bf r} \int\!d^3\!{\bf r^\prime} \biggl[
\rho_{p m_p;q m_q}({\bf r}) \, 
G({\bf r},{\bf r^\prime}&)& \,
\rho_{r m_r;s m_s}({\bf r^\prime}) \\       
~-~j^k_{p m_p;q m_q}({\bf r}&)& \, G^{kl}({\bf r},{\bf
r^\prime},|\epsilon_p-\epsilon_q|) \, j^l_{r m_r;s m_s}({\bf
r^\prime}) \biggr] \ , \nonumber
\end{eqnarray}
is expressed in terms of the scalar and tensor Green's functions,
$G({\bf r},{\bf r^\prime})$ and $G^{kl}({\bf r},{\bf r^\prime},
\omega)$, and the various components of the time-independent 
single-quark four-vector current
\begin{equation}
\overline{\psi}_{p m_p}({\bf r}) \, \gamma^\mu \, 
\psi_{q m_q}({\bf r})
~\equiv~
\left( \ \rho_{p m_p;q m_q}({\bf r}) \ , \ j^k_{p m_p;q m_q}({\bf r}) 
\ \right)
\ . 
\end{equation}
The first term under the integral in Eq. (35) stems from the
color-electrostatic interaction, which is responsible for color
confinement, and the second term arises due to the color-magnetic
hyperfine interaction, which yields the $N$-$\Delta$ splitting.

    As we restrict ourselves to values for the magnetic
single-particle quantum numbers of $m_p,m_q,\ldots \in \{ \pm 1/2
\}$, we can couple the $\hat z$-components of the angular momenta of
the two ``incoming" (or ``outgoing") quarks to two-body ``pseudospin"
\cite{St89} states, $|SM>$, where $|SM> \in \{ |00>, |10>, |1\pm 1>
\}$.  The corresponding formalism was outlined in detail in ref.
\cite{St89}.  This allows the separation of the gluonic one-body and
two-body contributions to the effective Hamiltonian into matrix
elements of three different spin-operators, ${\cal O}_0$, ${\cal
O}_z$ and ${\cal O}_\bot$, as was carried out in Eq. (33).  

\section{RESULTS AND DISCUSSION}

    In the following, we will present our results for the adiabatic,
local $N$-$N$ potential,
\begin{equation}
V_{ad}^{NN}~=~<H_1^{bag}>~+~<H_{OGE}>~-~2 \, E_{N}
\ ,  
\end{equation}
obtained in Born-Oppenheimer approximation \cite{BO} from the energy
difference of a deformed six-quark bag and two well separated
non-interacting nucleons.  The underlying effective Hamiltonian can
be separated into various contributions.  There is the one-body term,
\begin{eqnarray}
<H_1^{bag}>~=~
\epsilon_\sigma < & b_\sigma^+ & b_\sigma >~+~\epsilon_\pi 
< b_\pi^+ b_\pi >
\nonumber \\
-~&{1\over 2}&\int\!d^3\!{\bf r} \left( {dg_{eff}(\sigma) \over
d\sigma} \, \sigma <\overline\psi \psi> \, + \, {dU(\sigma) \over
d\sigma} \, \sigma \, - \, 2 U(\sigma)
\right) \\ 
&+&\,\int\!d^3\!{\bf r} \bigl( \, g_{eff}(\sigma) -
g_{eff}(\sigma_\alpha) \,
\bigr) <\overline\psi \psi> \nonumber
\end{eqnarray}
with the scalar quark density
\begin{equation}
<\overline\psi \psi>~\equiv~ <b_\sigma^+ b_\sigma >
\overline{\psi}_{\sigma m}\psi_{\sigma m} ~+~ <b_\pi^+ b_\pi >
\overline{\psi}_{\pi m}\psi_{\pi m}
\ .
\end{equation} 
$<H_1^{bag}>$ arises from the single-particle energies,
$\epsilon_\sigma$ and $\epsilon_\pi$ of Eq. (16) and depicted in Fig.
4, the scalar field $\sigma$ from Eq. (17) as well as the external
potential generating the quark wave functions,
$g_{eff}(\sigma_\alpha)$ of Eq. (18).  Note the distinction between
the scalar field $\sigma$, which is an explicit dynamical degree of
freedom, and the auxiliary quantity $\sigma_\alpha$, which, as
outlined in Sec. III, is only used for generating single-quark wave
functions with a certain deformation.

    In addition, there is the one gluon exchange contribution
\begin{eqnarray}
<H_{OGE}>~=~<H_1^{\sigma}>&+&<H_1^{\pi}>+
<H_2^{\sigma\sigma\sigma\sigma}>+ <H_2^{\pi\pi\pi\pi}> \nonumber \\
&+&<H_2^{\sigma\sigma\pi\pi}>+<H_2^{\sigma\pi\pi\sigma}>+
<H_2^{\sigma\pi\sigma\pi}> \ ,
\end{eqnarray}
with the one-body self-energy terms, $<H_1^{\sigma}>$ and
$<H_1^{\pi}>$, and the various two-body contributions $<H_2^{pqrs}>$
allowed by parity conservation.  The relevant expressions can be
derived from Eqs. (33) through (36), and the corresponding diagrams
are shown in Fig. 5.  It is important to note that in the OGE
self-energy terms only the color-electrostatic interaction was taken
into account, and that the $\sigma-\pi-\sigma$ and $\pi-\sigma-\pi$
``off-diagonal" self-energy terms are essential in providing for the
confinement of the color-electric flux \cite{DT77}.

    The adiabatic potential of Eq. (37) will be shown as a function
of the deformation parameter $\alpha$, which was introduced in Sec.
III (see Eqs. (19) to (21)).  For large prolate deformations,
$\alpha$ coincides with the true nucleon-nucleon separation.  For
smaller deformations, however, only a dynamical calculation employing
e.g. the generator coordinate method would yield the transformation
to the exact inter-nucleon separation \cite{Schuh}.  In Fig. 6 we
show the dependence of the inter-nucleon separation $r$ on the
deformation parameter $\alpha$ as taken from ref.
\cite{Schuh} where the $N$-$N$ interaction was investigated in terms
of quark degrees of freedom within the Friedberg-Lee soliton model
and not including gluonic effects.  We note that the``spherical"
configuration, $\alpha=0$, corresponds to a still finite
inter-nucleon separation, and that $r \to 0$ is approached for oblate
deformations, i.e.  for $\alpha<0$.  Although the exact form of the
transformation $r(\alpha)$ depends on the details of the Hamiltonian
and can thus only be established by a consideration of the dynamics
involved, which we leave to subsequent work \cite{Fut}, the general
behavior will still be similar to the one depicted in Fig. 6.

    Approximate recoil corrections \cite{recoil}, momentum projection
and boost \cite{boost} are not incorporated into the present
six-quark calculations.  For consistency, the potentials we calculate
here are therefore normalized with respect to the energy and not the 
mass of two non-interacting nucleons.  We find this quantity from a
self-consistent calculation of the nucleon, as outlined in Sec. II.
The latter also allows the adjustment of the parameters of the
model, which are fit to the experimental proton rms charge radius
(0.83 fm), the $N$-$\Delta$ mass splitting (293 MeV) and the nucleon 
mass (939 MeV),
\begin{equation}
M_N~=~\sqrt{E_N^2 \, - \, <P^2>}
\ , 
\end{equation}
where the approximate recoil corrections $<P^2>$ stem from the quark
wave functions and from the scalar $\sigma$-field.  Results of these
calculations are summarized in Table I, and the two parameter sets 
for which the $N$-$N$ interaction was evaluated are given in Table 
II.  The quantity $E_N$ varies between 1145 MeV for the set with
$f=\infty$ and 1212 MeV for $f=3$.

    Throughout this investigation, the strong coupling constant
$\alpha_s$ we use for evaluating the gluonic share of the effective
Hamiltonian is obtained by fitting the experimental $N$-$\Delta$
splitting employing either a ``free" or a ``confined" gluonic
propagator.  The corresponding values for $\alpha_s^{free}$ and
$\alpha_s^{conf}$ for the two parameter sets under consideration are
listed in the sixth and seventh column of Table II, respectively.
Thus, when evaluating the color-magnetic hyperfine interaction,
where, as pointed out in the last section, a ``free" gluonic
propagator is used, $\alpha_s^{free}$ is substituted for the strong 
coupling constant.  Correspondingly, $\alpha_s^{conf}$ is used for
the color-electrostatic part of the interaction where, on the other
hand, ``confined" propagators are employed.

    In addition, in all actual numerical calculations an infrared
regularization \cite{F88} of the dielectric function,
$\kappa(\sigma)$ of Eq. (7), was introduced in order to handle the 
infinities in the one gluon exchange diagrams associated with a 
vanishing dielectric constant.  We replace $\kappa(\sigma)$ with
\begin{equation}
\kappa(\sigma)~\rightarrow~\kappa(\sigma) \, (1-\kappa_v) ~+~ 
\kappa_v
\end{equation}
and use $\kappa_v=0.1$ throughout this investigation after having
convinced ourselves that our final results are stable with respect to
variations in the regularization parameter $\kappa_v$.  Note that
``intermediate quantities" -- as e.g. the energies associated with
individual diagrams or the strong coupling constant $\alpha_s$ --
will, however, well depend on $\kappa_v$.

    In Figs. 7 and 8, we show the adiabatic $N$-$N$ potential
obtained from a diagonalization of the effective Hamiltonian in the
Hilbert space spanned by the six-quark configurations listed in Eq.
(25).  Results are depicted for the isospin-spin channels $(TS)=(01)$
(Fig. 7) and $(TS)=(10)$ (Fig. 8).  For $(TS)=(01)$, the potential
can furthermore be split into a central and a tensor part, where
\begin{mathletters}
\begin{eqnarray}
V_{cent}^{(TS)=(01)} ~ &=&~ {1 \over 3}
\left( \, 2 V_{M=\pm 1}^{(TS)=(01)} \, 
+ \, V_{M=0}^{(TS)=(01)} \right)
\ , \\       
V_{tens}^{(TS)=(01)} ~ &=&~ {1 \over 6}
\left( \, \phantom{2} V_{M=\pm 1}^{(TS)=(01)} \, 
- \, V_{M=0}^{(TS)=(01)} \right)
\ . 
\end{eqnarray}
\end{mathletters}

    The central interaction we find is purely repulsive with a
``soft" core whose maximum varies between 200 MeV ($f=3$, solid line)
and 350 MeV ($f=\infty$, dot-dashed line) for the two parameter sets
under consideration.  As outlined in Sec. II, for $f=3$ the bag
pressure vanishes which leads to hard bags with a thin surface, and
for $f=\infty$ $U(\sigma=0)$ turns from a second minimum to an
inflection point which, in turn, generates soft bags with a thick
surface.  This leads to a very different behavior of the surface
energy associated with the scalar $\sigma$-field which extends to the
variations observed in the interaction.  Also, the cusp in the
central potential around $\alpha \approx 2.3$ fm for $f=3$ originates
in those particular surface dynamics.  At that deformation the scalar
field turns abruptly from forming two separated bags to spanning just
one united cavity.

    An intermediate range attraction, as observed in earlier
calculations of that type (see e.g. ref. \cite{DT77}) is not at all
visible in our results.  De Tar \cite{DT77} attributes the latter to
the strong color-electrostatic attraction within the quark triplets.
Although the color-electrostatic one gluon exchange diagrams are
entirely attractive, in our investigation their effects are actually
more than cancelled by the repulsive self-energy diagrams,
$<H_1^{\sigma}>$ and $<H_1^{\pi}>$.  Note that the color-magnetic
self-energies were left out altogether in this investigation.  On the
other hand, the long and medium range $N$-$N$ attraction should
actually be attributed to the meson exchange, and to get a good
description of this in a quark model presumably requires the ``sea"
quarks to be taken into account explicitly \cite{H16}, which are not
accounted for in our investigation at this stage.  To cure that
shortcoming, we plan \cite{Pion} to include an explicit pion exchange
between the quarks \cite{CT75} which will then lead to an effective
pionic dressing of the individual nucleons along the lines proposed
by Miller et al.
\cite{CBM}.

    To determine the relevance of the so-called ``hidden-color"
\cite{H81} states ($|3>$ through $|7>$ in Eq. (25)), which
asymptotically fission into color non-singlets,
in Fig. 9 we show their relative admixture to the six-quark ground
state of the effective Hamiltonian as a function of the deformation
of the bag.  We observe that their contributions become significant
as soon as the nucleonic bags overlap considerably, and that up to 50
percent of the ground state wave function can actually be made up of
``hidden-color" states for small inter-nucleon separations, i.e.
oblate deformations.  This proves the importance of channel coupling
in that realm, and is consistent with the findings of ref.
\cite{St89} as corrected in ref. \cite{Corr}.

    The different contributions to the adiabatic $N$-$N$ potential
are analyzed in Figs. 10 and 11 and in Table III.  In Fig. 10, we
show the various potentials we obtain when employing different
approximations for the one gluon exchange.  Results are depicted for
the isospin-spin channel $(TS)=(01)$ and for a two-nucleon state with
the spins aligned antiparallel along the separation axis, i.e. for
$M=0$.

    The dashed line corresponds to a calculation where the OGE was
left out altogether, and in agreement with an earlier investigation
\cite{Schuh} where the Friedberg-Lee soliton model was applied to
$N$-$N$ scattering we find a strongly attractive adiabatic potential.
The dotted line shows the results of a calculation where only the
color-magnetic hyperfine interaction was included, which in the
literature is quoted as being responsible for the short-range
repulsive core \cite{BCKS}. In contrast to that belief, the spin-spin
interaction reduces the attraction but does not yield any repulsion.
The dot-dashed and the solid line correspond to calculations where,
in addition, the color-electrostatic OGE was included, and they
indeed yield a repulsive core.  In contrast to the solid line, which
shows the results of a calculation employing a ``confined" scalar
Green's function, $G({\bf r},{\bf r^\prime})$, the dot-dashed line
corresponds to the use of a free gluonic propagator, $G_f({\bf
r},{\bf r^\prime})$, as outlined in Sec. V. The differences between
both calculations prove to be rather minute, which gives us
confidence, that the uncertainties we encounter from also evaluating
the color-magnetic hyperfine interaction with a free tensor
propagator are as well quite small.

    In Fig. 11, the adiabatic potential for the isospin-spin channel
$(TS)=(10)$ is split into a gluonic contribution, $<H_{OGE}>$ of Eq.
(40), and a part independent of the one gluon exchange, $<H_1^{bag}>$
of Eq. (38).  The non-gluonic contribution, which stems from the
single particle energies, the scalar $\sigma$-field and the external
potential used to generate the quark wave functions is always
attractive, while the OGE part of the Hamiltonian is purely
repulsive.  Also, the differences between the two parameter sets,
which we observed in the central adiabatic potentials shown in Figs.
7 and 8, almost entirely stem from $<H_1^{bag}>$, with the ``hard"
set ($f=3$, the solid line) being much more attractive.  As already
previously mentioned, the origin of that variations is the very
different behavior of the surface energy associated with the scalar
$\sigma$-field for the two parameter sets under consideration.  In
spite of that differences, their gluonic shares of the energy are
very similar.

    At this point, a more detailed comparison with De Tar's
\cite{DT77} pioneering study of the adiabatic two-nucleon interaction
in the framework of the MIT bag model is in place.  As can be seen
from Figs. 8 and 9 of ref. \cite{DT77}, for the MIT bag model the
non-gluonic contribution, $<H_1^{bag}>$, is repulsive, while the
gluonic share of the Hamiltonian, $<H_{OGE}>$, yields all the
attraction.  This is just opposite to our findings.  The difference
concerning the non-gluonic interaction is due to the very different
nature of the surface dynamics of the scalar background field for the
MIT and the solitonic bag.  The differences in the gluonic share of
the Hamiltonian, on the other hand, arise most probably both from the
approximations De Tar is making in the evaluation of the OGE
self-energy terms and from the differences in the color-dielectric
constant $\kappa({\bf r})$ and thus also in the gluonic propagators
between the two models.  Also, the six-quark configuration space we
are using is much larger than the one De Tar was employing, and his
single-quark states are rather artificial constructions while our
``molecular orbitals" are eigen-states of the constrained mean field 
Hamiltonian.

    To get a more detailed understanding of the origins of our
results, in Table III we list the various contributions to the
adiabatic $N$-$N$ potential stemming from the individual diagrams
shown in Fig. 5.  The energies shown are the expectation values of
the ground state of the effective Hamiltonian for the isospin-spin
channel $(TS)=(10)$.  They correspond to the limiting cases of two
well separated nucleons ($\alpha=3.5$ fm) as well as one united
spherically symmetric cavity ($\alpha=0.0$ fm).  The individual OGE
mutual interaction terms $<H_2^{pqrs}>$ are all attractive, while the
self-energy diagrams, $<H_1^{\sigma}>$ and $<H_1^{\pi}>$, are
entirely repulsive.  Furthermore, only if the color-electrostatic
``off-diagonal" self-energy terms $\sigma-\pi-\sigma$ and
$\pi-\sigma-\pi$ are taken into account, the color-electric flux is
confined \cite{DT77}, and the interaction vanishes at asymptotic
inter-nucleon separations, i.e. $<H> \to 0$ for large $\alpha$.  That
this can really be observed in our results -- see the forth and
seventh column in the table -- is non-trivial and is a nice
confirmation that our numerics are correct.  It can also be seen from
Table III, that the adiabatic potential at vanishing $N$-$N$
separation arises from significant cancellations between individual
terms that are rather large, with the repulsive one gluon exchange
contributions overpowering the attractive non-gluonic share of the
effective Hamiltonian.

    For all the results reported in this section, the single-quark
wave functions were generated from a scalar potential having a
particular shape, as characterized by the geometrical parameters
$R(\alpha)$, $\Gamma$ and $\sigma_0$ (see Eqs. (19) to (21)). As
outlined in Sec. III, the radius of the six-quark bag $R(\alpha)$ was
varied as a function of the deformation such that the volume of the
$\sigma$-field cavity remains fixed.  In order to test the validity
of that method, for $\alpha=0$ we constructed self-consistent
eigen-states of the total Hamiltonian requiring a spherically
symmetric scalar field.  In Table IV, we compare results for the
geometrical shape parameters and the adiabatic potential, which we
find from the self-consistent calculation with the corresponding
quantities obtained by using the ``constant volume" approach.
Results are listed for $(TS)=(01)$ and $M=0$.  Although the
geometrical parameters of the self-consistently determined
$\sigma$-field are quite different from the ones characterizing
$\sigma_\alpha$ in the ``constant volume" approximation, the resulting
adiabatic potentials are very similar.  This shows, that the
uncertainties we encounter by choosing this particular path in the
configuration space spanned by the geometrical parameters $R$,
$\Gamma$ and $\sigma_0$ will also be minute and should hence not
effect our conclusions.

\section{CONCLUSION AND OUTLOOK}

    We have evaluated the adiabatic nucleon-nucleon potential in a
relativistic quark bag model which yields spatial as well as color
confinement and is free of the spurious color Van der Waals forces
troubling most non-relativistic calculations in that realm. The
six-quark system we investigate is confined in a deformed bag-like
mean field through an effective non-linear interaction between the
quarks and a scalar field. The shape of this confining field is
adjusted to reproduce the corresponding quantity for the asymptotic
case of two well separated non-interacting nucleons, and is then
varied with deformation treating the scalar field as an
incompressible liquid.

     Six-quark molecular-type configurations are then generated as a
function of deformation and their energies are evaluated in a coupled
channel analysis. By using molecular states instead of cluster model
wave functions, we can be sure that all important six-quark
configurations were properly taken into account, which is a necessary
prerequisite for finding reasonable results.

    The corresponding effective Hamiltonian includes not only the
interaction between the quarks and the scalar background field but
also quark-quark interactions generated through one gluon exchange
evaluated in Abelian approximation. Furthermore, when calculating the
gluonic propagators mediating that interaction, the inhomogeneity and
deformation of the dielectric medium were taken into account and the
Coulomb gauge was applied.

    Results for the adiabatic local nucleon-nucleon potential have
been presented for the different spin-isospin channels which are
compatible with $L=0$ $N$-$N$ partial waves, and they differ quite
considerably from a realistic phenomenological interaction fit to the
experimental phase shifts.  Although the adiabatic central potentials
display a ``soft" repulsive core, as is desirable from
phenomenology, they totally lack the intermediate range attraction,
which was observed in earlier calculations of that type and which was
attributed to the strong color-electrostatic attraction \cite{DT77}.

    Although the color-electrostatic exchange diagrams are entirely
attractive also in our investigation, their effects are actually more
than cancelled by the repulsive gluonic self-energy diagrams. A
detailed analysis of the different contributions to the effective
Hamiltonian unveils that the non-gluonic one-body terms would lead to
considerable attraction for vanishing inter-nucleon separation, while
the one gluon exchange mutual and self-interaction terms produce all
the repulsion. To be more specific, in our case it is the
color-electric one gluon exchange which leads to the repulsion at
small $N$-$N$ separations and not the spin-spin color-magnetic
hyperfine interaction, which in the literature is quoted as being
responsible for the short-range repulsive core \cite{BCKS}.

    Considering that the long and medium range nucleon-nucleon
attraction should actually be attributed to explicit meson exchange
and not to quark rearrangement, we are not at all surprised not to
get a good description of this part of the interaction in a quark
model which does not include the ``sea" quarks. We plan to overcome
that detriment by either including quantum surface fluctuations,
which would introduce configurations of the form $q^7\bar q$ in
addition to our $q^6$ basis states, or by considering an explicit
pion exchange between the individual quarks along the lines followed
in the cloudy bag model \cite{CBM}.  The latter mechanism is
favorable as it also leads to a restoration of the broken chiral
symmetry.  Work in that direction is currently in progress
\cite{Pion}.

    Finally, we also plan to account for the dynamics of the $N$-$N$
interaction by extending this work through means of the generator
coordinate method \cite{Fut}. It has been shown that a significant
part of the short-range $N$-$N$ repulsion is due to dynamics
\cite{Schuh}, and that the absence of a repulsive core in some early
calculations was an artifact of the adiabatic approximation
\cite{F83,H84}.  In addition, the effective $N$-$N$ interaction is
highly non-local in terms of the separation parameter.

\acknowledgments{We wish to thank P. Tang and S. Hartmann for many
useful discussions.  This work was supported in part by the U.S.
Department of Energy, by the Deutsche Forschungsgemeinschaft (W.K.)
and by the Feodor-Lynen program of the Alexander von
Humboldt-Stiftung (W.K.).}

\begin{table}
\caption{Results of a self-consistent mean field calculation for the
nucleon for various parameter sets as characterized by $f=b^2/ac$ and
$c$, and adjusted to yield the same recoil corrected proton rms
radius of 0.83 fm and recoil corrected proton mass of $939$ MeV. The
quantities listed are the $0^{++}$ glueball mass $m_{GB}$, the bag
pressure $B$, the nucleon's axial vector coupling constant $g_A$ and
the proton's magnetic moment $\mu_p$.  The strong coupling constant
$\alpha_s$ can either be adjusted to yield the mass of the
$\Delta$-resonance  of $1232$ MeV -- employing a free or a
``confined" gluonic propagator -- or can be obtained from an
evaluation of the string tension, $\theta$.  Corresponding values for
$\alpha_s$ are listed.  For further details see ref.
\protect\cite{Book} and references therin.}
\medskip
\begin{tabular}{dd@{  }|ddddddd}
$f$ & $c$ & $m_{GB}$ (MeV) & $B$ (MeV/fm$^3$) & $g_A$ & $\mu_p$
($\mu_N$) & \multicolumn{2}{c}{$\alpha_s|_{N\Delta}$}   
& $\alpha_s|_{\theta}$ \\ & & & &
& & free & confined & \\
\hline
3.0     & 30000 & 2933 &  0 & 1.24 & 2.21 & 3.36 & 1.32 & 2.93 \\ &
10000 & 1948 &  0 & 1.21 & 2.29 & 3.36 & 1.42 & 3.78 \\ &  3000 &
1254 &  0 & 1.18 & 2.34 & 3.34 & 1.57 & 4.78 \\ &  1000 &  734 &  0 &
1.12 & 2.42 & 2.90 & 2.25 & 13.20 \\ 3.2     & 30000 & 2501 & 40 &
1.21 & 2.28 & 3.49 & 1.43 & 1.83 \\ & 10000 & 1787 & 32 & 1.20 & 2.31
& 3.48 & 1.46 & 2.08 \\ &  3000 & 1214 & 22 & 1.19 & 2.34 & 3.45 &
1.56 & 2.53 \\ &  1000 &  783 & 12 & 1.15 & 2.38 & 3.18 & 2.03 & 4.47
\\ $\infty$& 30000 & 2355 & 67 & 1.21 & 2.31 & 3.66 & 1.47 & 1.61 \\
& 10000 & 1755 & 62 & 1.21 & 2.31 & 3.64 & 1.50 & 1.67 \\ &  3000 &
1243 & 52 & 1.20 & 2.32 & 3.58 & 1.59 & 1.88 \\ &  1000 &  874 & 38 &
1.18 & 2.35 & 3.49 & 1.79 & 2.43 \\
\end{tabular}
\end{table}

\begin{table}
\caption{Parameter sets for which the adiabatic $N$-$N$ potential was
calculated. The sets are adjusted to yield the proton rms charge
radius, the nucleon mass and the $N$-$\Delta$ mass splitting, where
the latter quantity was evaluated employing a free as well as a
``confined" gluonic propagator. For further details see Sec. II.}
\medskip
\begin{tabular}{ddddddd}
$f$ & $a$ (fm$^{-2}$) & $b$ (fm$^{-1}$) & $c$ & $g_0$ &
\multicolumn{2}{c}{$\alpha_s|_{N\Delta}$}\\ 
&&&&& $\alpha_s^{free}$ & $\alpha_s^{conf}$ \\
\hline
3.0      & 97.45 & -1709.9 & 10000 & 0.81 & 3.36 & 1.42 \\ $\infty$ &
~0.00 &  -726.1 & 10000 & 1.80 & 3.64 & 1.50 \\
\end{tabular}
\end{table}

\begin{table}
\caption{The various contributions to the adiabatic $N$-$N$ potential
stemming from the individual diagrams shown in Fig. 5 and given in
Eqs. (38) and (40).  Results are shown for well separated
($\alpha=3.5$ fm) as well as completely overlapping nucleonic bags
($\alpha=0.0$ fm), and also for the adiabatic potential at
$\alpha=0.0$ fm, which is the difference of the latter two
quantities. The energies shown correspond to the isospin-spin channel
$(TS)=(10)$ and are obtained using the parameter sets given in Table
II.}
\medskip
\begin{tabular}{c@{\quad}|rrr@{\quad}|rrr}
& \multicolumn{3}{c|}{$f=3.0$} & \multicolumn{3}{c}{$f=\infty$} \\ 
& $E|_{0.0\,\hbox{\scriptsize fm}}$ 
& $E|_{3.5\,\hbox{\scriptsize fm}}$ 
& $V_{ad}^{NN}|_{0.0\,\hbox{\scriptsize fm}}$ 
& $E|_{0.0\,\hbox{\scriptsize fm}}$ 
& $E|_{3.5\,\hbox{\scriptsize fm}}$
& $V_{ad}^{NN}|_{0.0\,\hbox{\scriptsize fm}}$ 
\\
\hline
$<H_1^{bag}>$ 
& 93 & 240 & $-$147 & 212 & 256 & $-$44 \\
\hline
$<H_1^{\sigma}>$ 
& 5849 & 3757 & 2092 & 5506 & 4685 & 821 \\
$<H_1^{\pi}>$ 
& 2654 & 3756 & $-$1102 & 3093 & 4682 & $-$1589 \\
\hline
$<H_2^{\sigma\sigma\sigma\sigma}>$
& $-$3728 & $-$1100 & $-$2628 & $-$3222 & $-$1507 & $-$1715 \\ 
$<H_2^{\pi\pi\pi\pi}>$ 
& $-$901 & $-$1098 & 197 & $-$1101 & $-$1504 & 403 \\
$<H_2^{\sigma\sigma\pi\pi}>$ 
& $-$3403 & $-$2504 & $-$899 & $-$3651 & $-$3390 & $-$261 \\ 
$<H_2^{\sigma\pi\pi\sigma}>$ 
& $-$92 & $-$1402 & 1310 & $-$144 & $-$1480 & 1336 \\ 
$<H_2^{\sigma\pi\sigma\pi}>$ 
& $-$403 & $-$1649 & 1246 & $-$518 & $-$1742 & 1224 \\
\hline
$<H>$ 
& 69 & 0 & 69 & 175 & 0 & 175 \\
\end{tabular}
\end{table}

\begin{table}
\caption{The geometrical parameters, $R(\alpha)$, $\Gamma$ and
$\sigma_0$, characterizing the field $\sigma_\alpha$ (see Eqs. (19)
to (21)) from which then the external scalar potential,
${\cal{V}}_\alpha$ of Eq. (18), is obtained.  The quantities shown
correspond to the isospin-spin channel $(TS)=(01)|_{M=0}$ and to zero
deformation, i.e. $\alpha = 0$ fm. Self-consistent solutions
constrained to yield a spherically symmetric scalar field are
compared with the ``constant volume" approach outlined in Sec. III.}
\medskip
\begin{tabular}{d@{  }|cdddd}
f &&$R$ (fm)&$\Gamma$ (fm$^{-1}$)&$\sigma_0$
(fm$^{-1}$)&$V_{ad}^{NN}$ (MeV)\\
\hline
3.0      & constant volume & 1.021 & 4.230 & 0.4050 & 128.8 \\ &
self-consistent & 1.178 & 5.548 & 0.3514 & 124.2 \\
\hline
$\infty$ & constant volume & 1.226 & 2.167 & 0.2592 & 264.5 \\ &
self-consistent & 1.257 & 2.032 & 0.2249 & 262.3 \\
\end{tabular}
\end{table}

\begin{figure}
\ifnum\figcond>0 
\vbox to 6.30in{\centerline{\hbox{\hsize=6in\hss\vbox to 6in{\vss
\centerline{\epsfbox{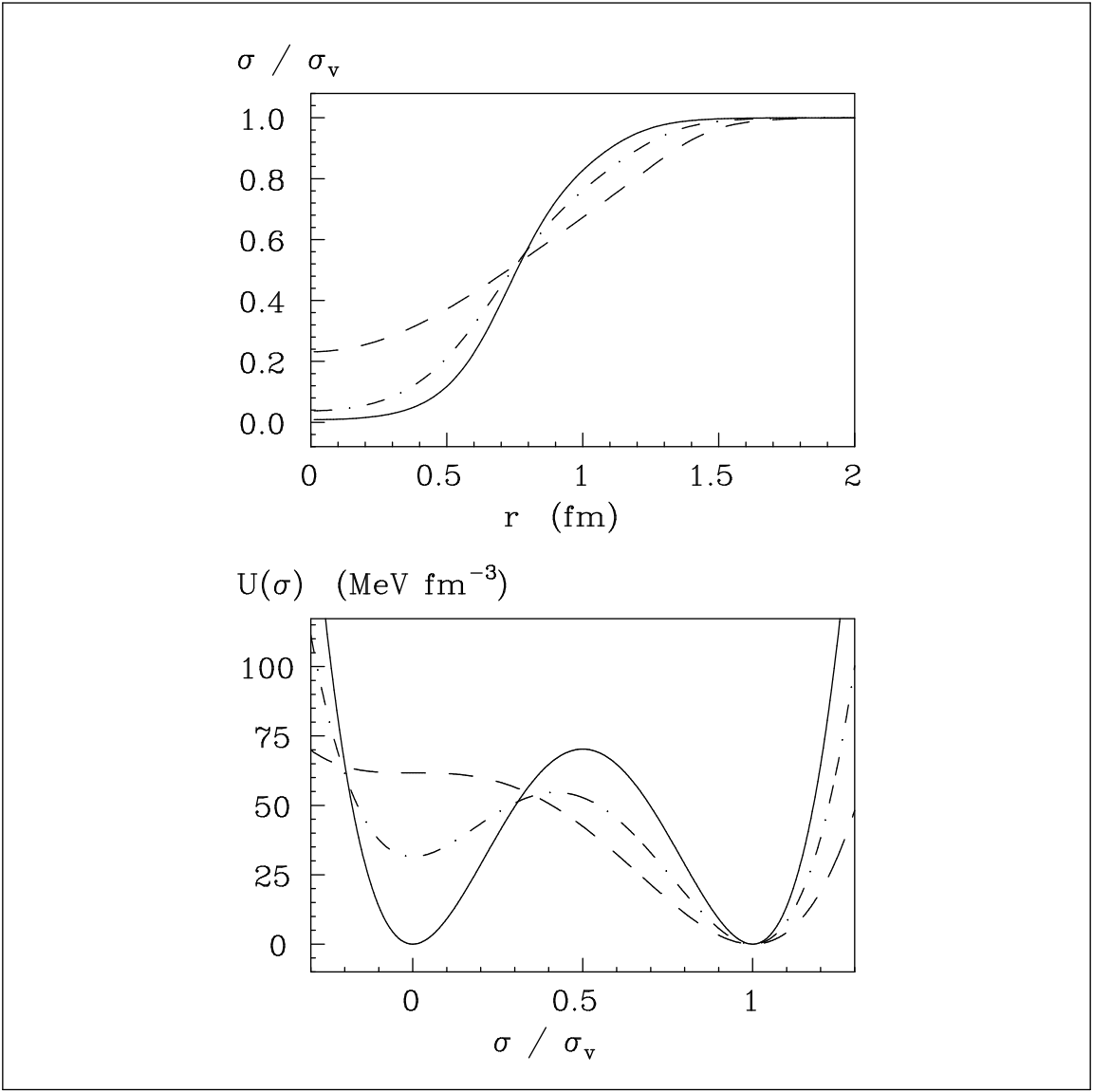}}\vss}\hss}}\vfill}
\fi
\caption{The scalar field, $\sigma/\sigma_v$, and the corresponding
self-interaction potential, $U(\sigma)$ of Eq. (4), determined from a
self-consistent evaluation of the nucleon for three different
parameter sets.  The solid line corresponds to $f=3$, the dot-dashed
line to $f=3.2$ and the dashed line to $f=\infty$.  For all sets $c$,
is kept at a value of $10000$.}
\end{figure}

\begin{figure}
\ifnum\figcond>0 
\vbox to 4.00in{\centerline{\hbox{\hsize=6in\hss\vbox to 3.7in{\vss
\centerline{\epsfbox{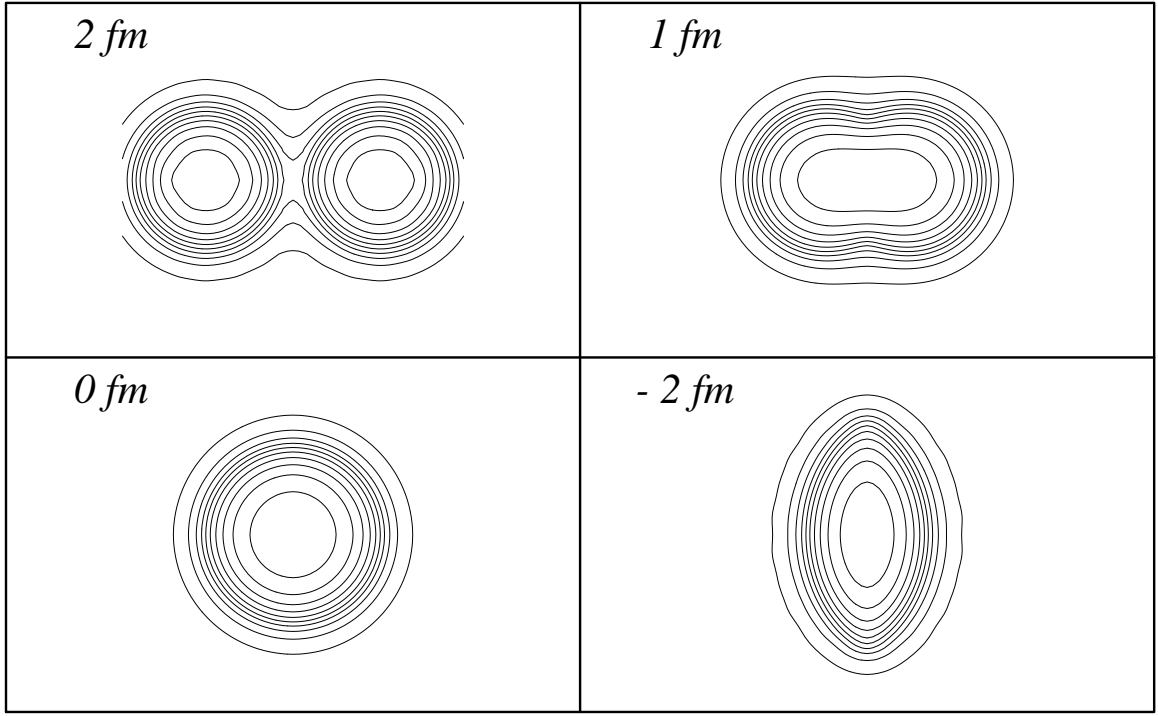}}\vss}\hss}}\vfill}
\fi
\caption{The scalar field, $\sigma_\alpha$ of Eq. (19), from which
the single-quark wave functions are generated by means of Eqs. (16)
and (18) for four different values of the deformation parameter
$\alpha$ between $2$ fm and $-2$ fm.  The fields correspond to the
parameter set with $f=3$ and $c=10000$, and are shown with equal
increments between adjacent contours.}
\end{figure}

\begin{figure}
\ifnum\figcond>0 
\newpage
\vbox to 3.30in{\centerline{\hbox{\hsize=6in\hss\vbox to 3in{\vss
\centerline{\epsfbox{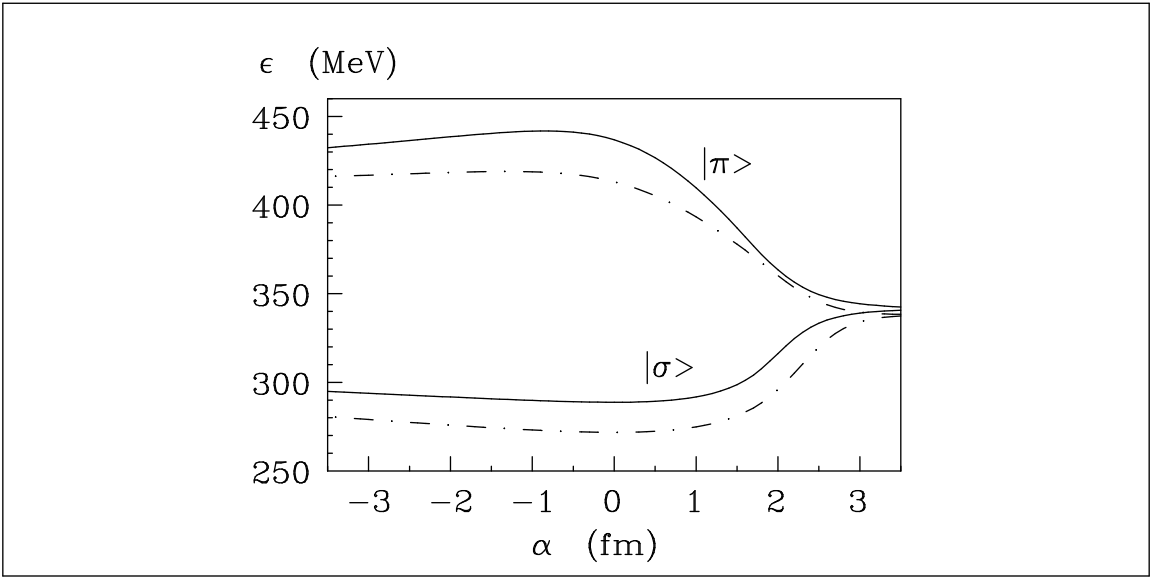}}\vss}\hss}}\vfill}
\fi
\caption{The eigenenergies of the lowest single-particle states of
positive and negative parity determined from Eq. (16) for values of
the magnetic quantum number of $m=\pm 1/2$. Results are shown for two
particular parameter sets with $c=10000$ and either $f=3$ (solid
line) or $f=\infty$ (dot-dashed line).}
\end{figure}

\begin{figure}
\ifnum\figcond>0 
\vbox to 3.30in{\centerline{\hbox{\hsize=6in\hss\vbox to 3in{\vss
\centerline{\epsfbox{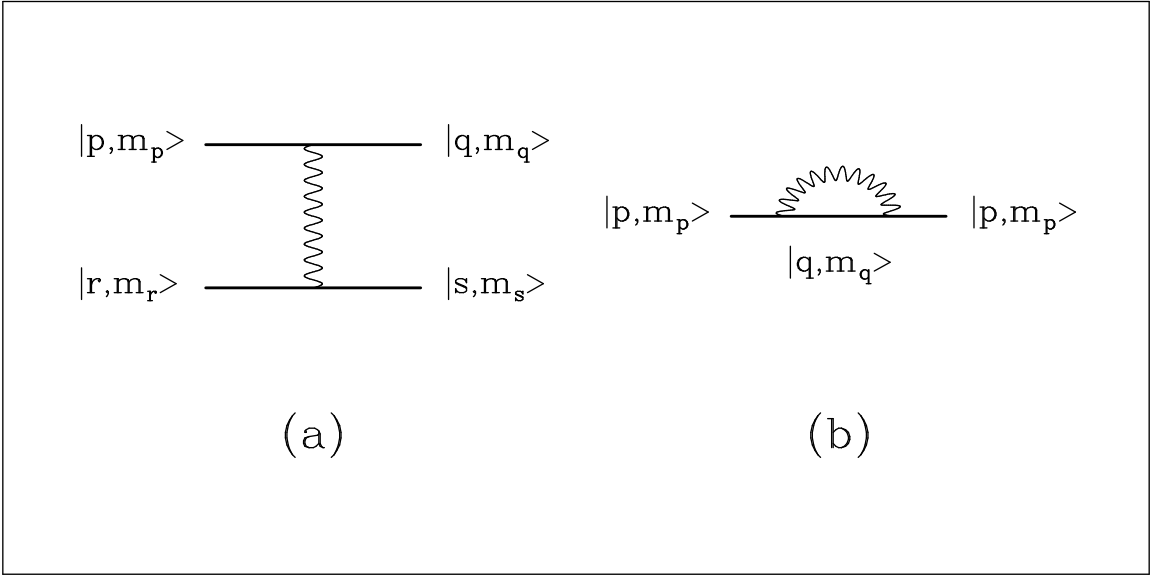}}\vss}\hss}}\vfill}
\fi
\caption{Typical one gluon exchange diagrams that contribute to the
gluonic share of the effective Hamiltonian, $<H_{OGE}>$ of Eq. (33).
The diagrams shown in (a) correspond to the two-body mutual
interactions, while the graphs depicted in (b) correspond to one-body
self-interactions.}
\end{figure}

\begin{figure}
\ifnum\figcond>0 
\vbox to 3.30in{\centerline{\hbox{\hsize=6in\hss\vbox to 3in{\vss
\centerline{\epsfbox{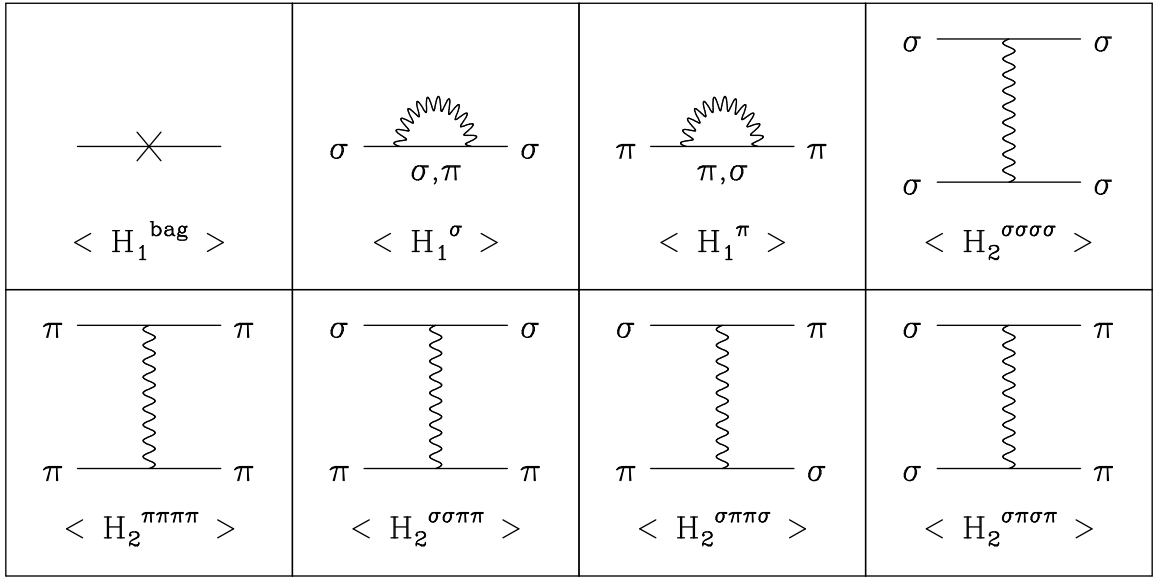}}\vss}\hss}}\vfill}
\fi
\caption{Diagrammatic representation of the various one-body
($<H_1^{\cdots}>$) and two-body ($<H_2^{\cdots}>$) contributions to
the effective Hamiltonian.}
\end{figure}

\begin{figure}
\ifnum\figcond>0 
\vbox to 3.30in{\centerline{\hbox{\hsize=6in\hss\vbox to 3in{\vss
\centerline{\epsfbox{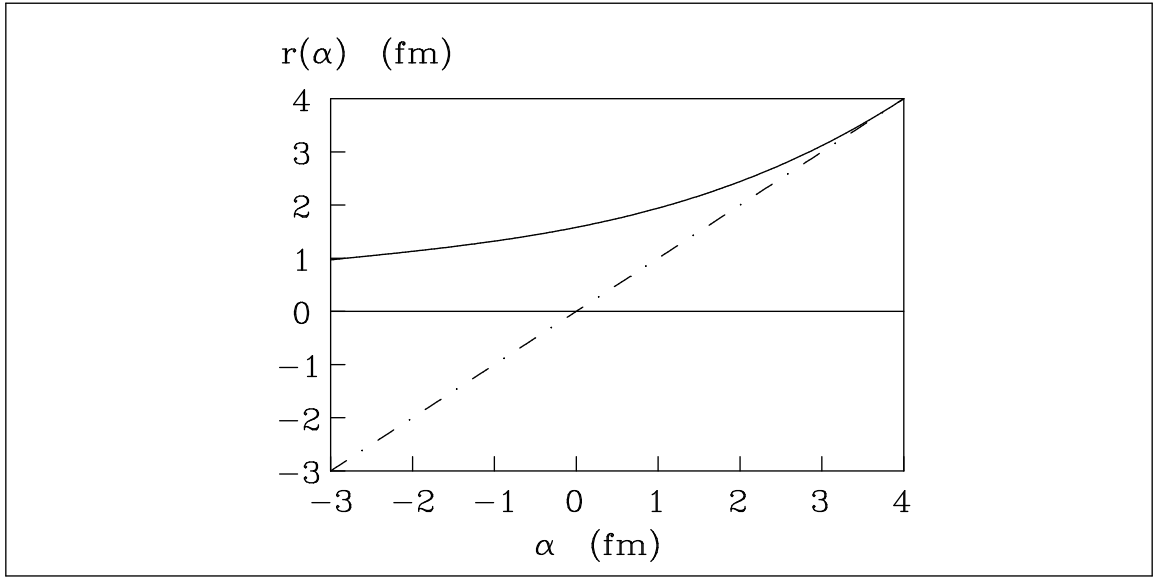}}\vss}\hss}}\vfill}
\fi
\caption{The inter-nucleon separation $r$ as a function of the
deformation parameter $\alpha$. The figure is taken from ref.
\protect\cite{Schuh} where the Friedberg-Lee soliton model was
applied to $N$-$N$ scattering.  No gluonic effects were taken into
account in ref. \protect\cite{Schuh}.}
\end{figure}

\begin{figure}
\ifnum\figcond>0 
\vbox to 6.30in{\centerline{\hbox{\hsize=6in\hss\vbox to 6in{\vss
\centerline{\epsfbox{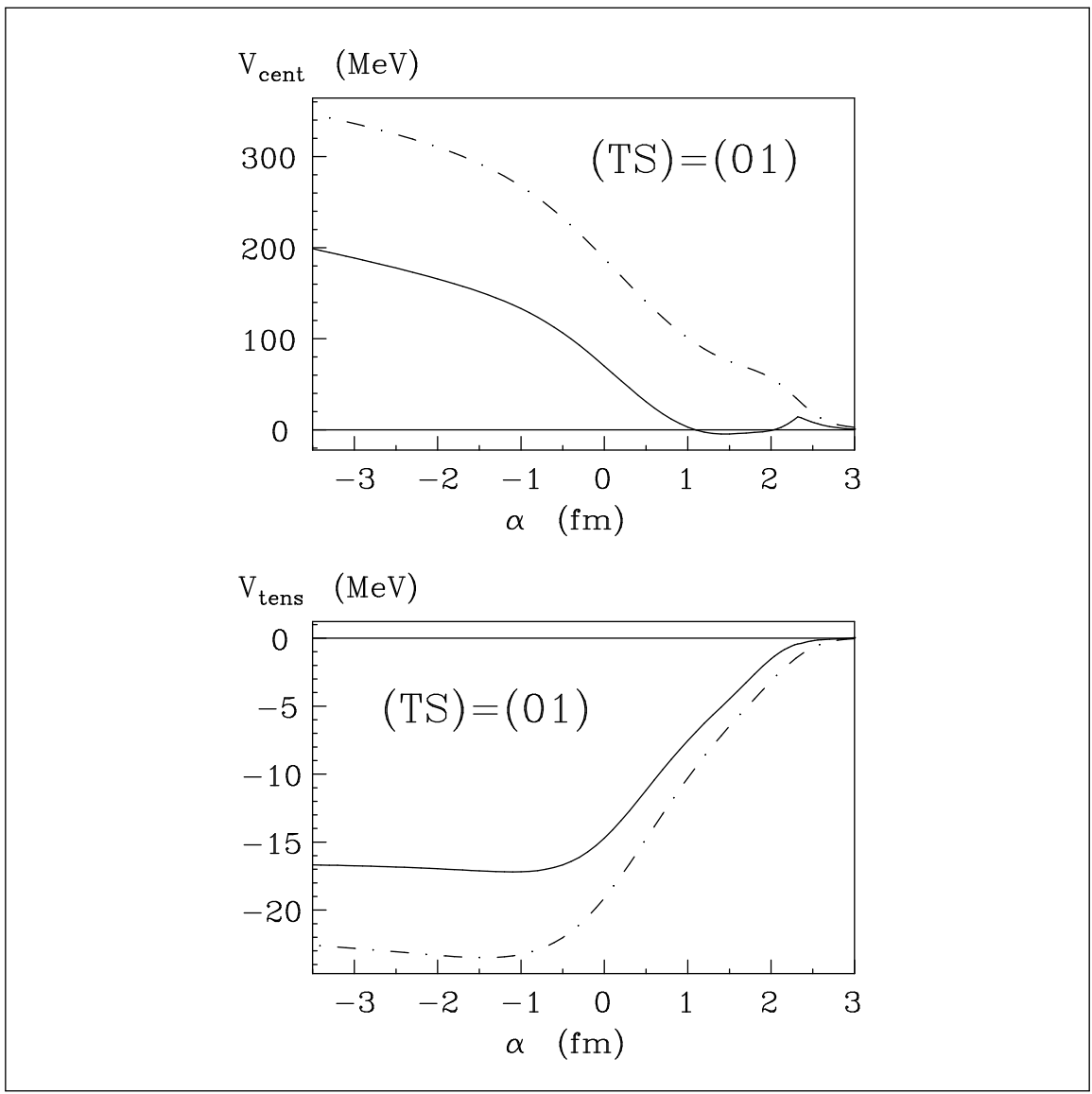}}\vss}\hss}}\vfill}
\fi
\caption{The adiabatic $N$-$N$ potential for the isospin-spin channel
$(TS)=(01)$ split into a central -- Eq. (43a) -- and a tensor part --
Eq.  (43b). The solid line corresponds to the parameter set with
$f=3$ and the dot-dashed line to the set with $f=\infty$.  Both sets
were adjusted to the standard properties of the nucleon and are
listed in Table II.}
\end{figure}

\begin{figure}
\ifnum\figcond>0 
\vbox to 3.30in{\centerline{\hbox{\hsize=6in\hss\vbox to 3in{\vss
\centerline{\epsfbox{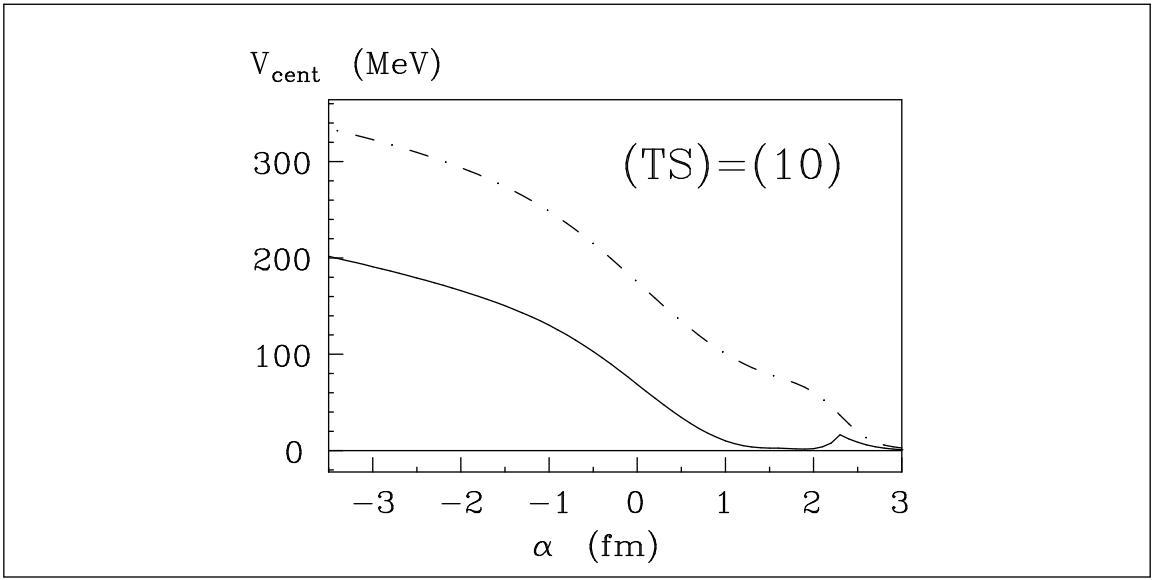}}\vss}\hss}}\vfill}
\fi
\caption{The adiabatic $N$-$N$ potential for the isospin-spin channel
$(TS)=(10)$. The labeling is the same as in Fig. 7.}
\end{figure}

\begin{figure}
\ifnum\figcond>0 
\vbox to 6.30in{\centerline{\hbox{\hsize=6in\hss\vbox to 6in{\vss
\centerline{\epsfbox{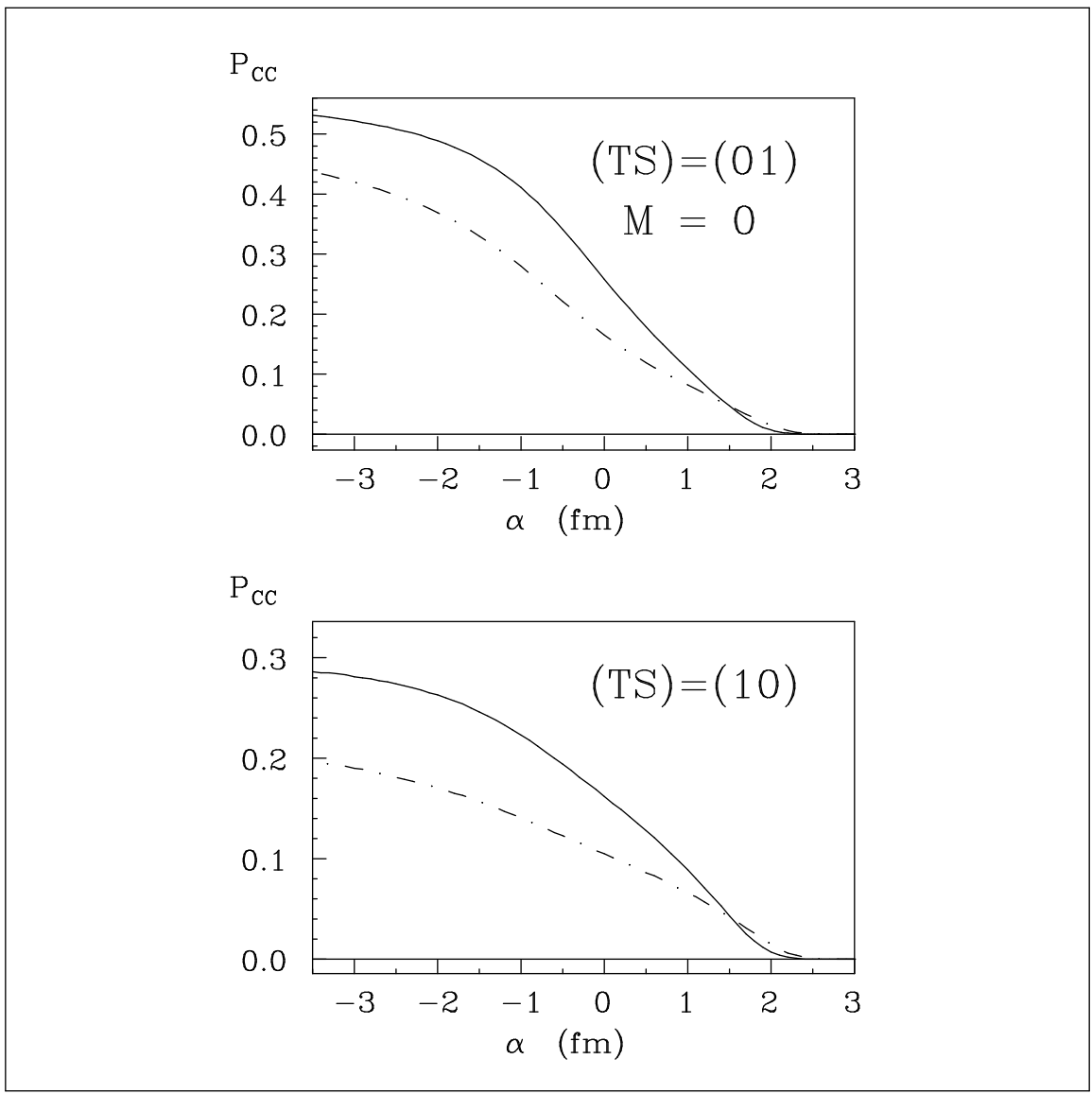}}\vss}\hss}}\vfill}
\fi
\caption{The relative admixture of ``hidden color" states, $|3>$
through $|7>$ in Eq. (25), to the ground state of the effective
Hamiltonian for two different isospin-spin channels,
$(TS)=(01)|_{M=0}$ and $(TS)=(10)$, and for the two parameter sets
listed in Table II.  The solid line corresponds to $f=3$ and the
dot-dashed line to $f=\infty$.}
\end{figure}

\begin{figure}
\ifnum\figcond>0 
\vbox to 6.30in{\centerline{\hbox{\hsize=6in\hss\vbox to 6in{\vss
\centerline{\epsfbox{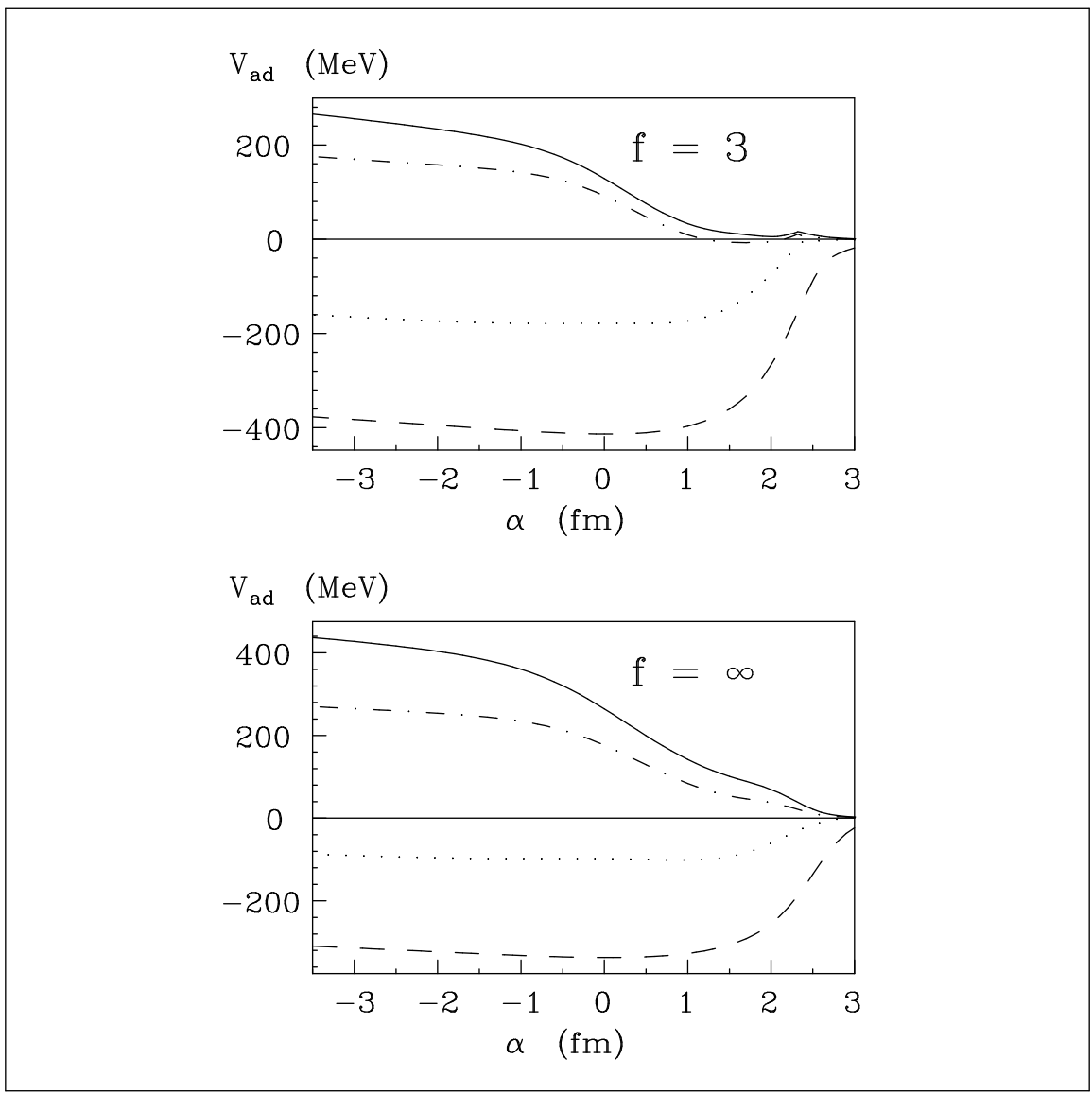}}\vss}\hss}}\vfill}
\fi
\caption{The various adiabatic $N$-$N$ potentials obtained when
employing different approximations for the one gluon exchange.
Results are shown for the isospin-spin channel $(TS)=(01)|_{M=0}$ and
for the two parameter sets given in Table II.  The dashed line
corresponds to a calculation where the OGE was left out altogether
and the dotted line shows the results of a calculation where only the
color-magnetic hyperfine interaction was included.  The dot-dashed
and the solid line correspond to calculations where, in addition,
different versions of the color-electrostatic OGE were taken into
account. In detail, the solid line shows the results of a calculation
employing a ``confined" Green's function, while the dot-dashed line
corresponds to the use of a free gluonic propagator.}
\end{figure}

\begin{figure}
\ifnum\figcond>0 
\vbox to 6.30in{\centerline{\hbox{\hsize=6in\hss\vbox to 6in{\vss
\centerline{\epsfbox{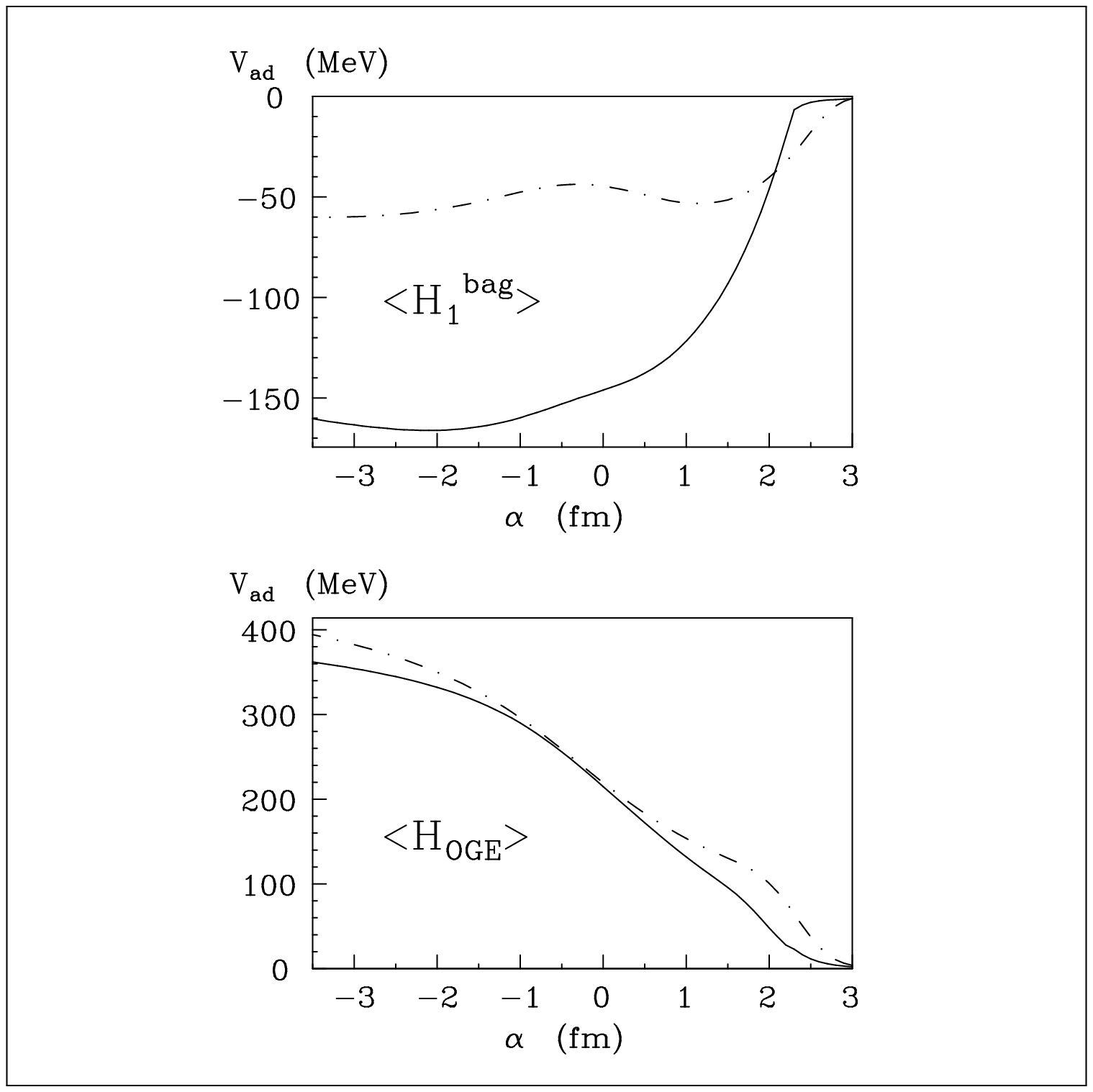}}\vss}\hss}}\vfill}
\fi
\caption{The adiabatic potential for the isospin-spin channel
$(TS)=(10)$ is split into a part independent of the one gluon
exchange, $<H_1^{bag}>$ of Eq.  (38), and a gluonic contribution,
$<H_{OGE}>$ of Eq. (40).  The solid line corresponds to $f=3$ and the
dot-dashed line to $f=\infty$.}
\end{figure}


\begin{references}
\bibitem{Y35}
 H. Yukawa, Proc. Phys.-Math. Soc. Japan {\bf 17}, 48 (1935).

\bibitem{M89}
 R. Machleidt, Adv. Nucl. Phys. {\bf 19}, 189 (1989).

\bibitem{Latt}
 H. Markum, M. Meinhart, G. Eder, M. Faber and H. Leeb, Phys. Rev. D
{\bf 31}, 2029 (1985).

\bibitem{L77}
 D.A. Libermann, Phys. Rev. D {\bf 16}, 1542 (1977).

\bibitem{DT77}
 C.E. De Tar, Phys. Rev. D {\bf 17}, 323 (1977).

\bibitem{Schuh}
 A. Schuh, H.J. Pirner and L. Wilets, Phys. Lett. {\bf 174B}, 10
(1986).

\bibitem{W91}
 T.S. Walhout and J. Wambach, Phys. Rev. Lett. {\bf 67}, 314 (1991).

\bibitem{W92}
 N.R. Walet, R.D. Amado and A, Hosaka, Phys. Rev. Lett. {\bf 68}, 
3849 (1992). 

\bibitem{O84} M. Oka and K. Yazaki, {\em Quarks and nuclei},
International review of Nuclear Physics, vol.1, ed. W. Weise (World
Scientific, Singapore, 1984) p. 489.

\bibitem{M88}
 F. Myhrer and J. Wroldsen, Rev. Mod. Phys. 60, 629 (1988).

\bibitem{S89}
 K. Shimizu, Rep. Prog. Phys. 52, 1 (1989).
  
\bibitem{BO}
 M. Born and J.R. Oppenheimer, Ann. Phys. {\bf 84}, 457 (1927).

\bibitem{O81}
 M. Oka and K. Yazaki, Prog. Theor. Phys. {\bf 66}, 556 (1981); {\em
ibid.} 572.
  
\bibitem{C83}
 M. Cvetic, B. Golli, N. Mankoc-Borstnik and M. Rosina, Nucl. Phys.
{\bf A395}, 349 (1983).

\bibitem{F88}
 G. Fai, R.J. Perry and L. Wilets, Phys. Lett. {\bf 208B}, 1 (1988).

\bibitem{K88}
 G. Krein, P. Tang, L. Wilets and A.G. Williams, Phys. Lett.  {\bf
212B}, 362 (1988); Nucl. Phys. {\bf A523}, 548 (1991).

\bibitem{VdW}
 R.S. Willey, Phys. Rev. D {\bf 18}, 270 (1978); P.M. Fishbane and
M.T. Gisaru, Phys. Lett. {\bf 74B}, 98 (1978); S. Matsuyama and H.
Miyazawa, Prog. Theor. Phys. {\bf 61}, 942 (1979); O.W. Greenberg and
H.J. Lipkin, Nucl. Phys. {\bf A370}, 349 (1981).

\bibitem{St87}
 Fl. Stancu and L. Wilets, Phys. Rev. C {\bf 36}, 726 (1987).

\bibitem{St88}
 Fl. Stancu and L. Wilets, Phys. Rev. C {\bf 38}, 1145 (1988).

\bibitem{St89}
 Fl. Stancu and L. Wilets, Phys. Rev. C {\bf 40}, 1901 (1989).

\bibitem{GCM}
 D.L. Hill and J.A. Wheeler, Phys. Rev. {\bf 89}, 1102 (1953); J.J.
Griffin and J.A. Wheeler, Phys. Rev. {\bf 108}, 311 (1957).

\bibitem{F83}
 A. Faessler, F. Fernandez, G. L\"ubeck and K. Shimizu, Nucl. Phys.
{\bf A402}, 555 (1983).

\bibitem{H84}
 M. Harvey, J. LeTourneux and B. Lorazo, Nucl. Phys. {\bf A424}, 428
(1984).

\bibitem{Fut}
 S. Pepin, W. Koepf, L. Wilets and Fl. Stancu, work in progress.

\bibitem{Tang}
 L. Wilets, S. Hartmann and P. Tang, {\em The Chiral Chromodielectric
Model: Quark Self Energy and Hadron Bags}, University of Washington
preprint (1993).

\bibitem{F77}
 R. Friedberg and T.D. Lee, Phys. Rev. D {\bf 15}, 1694 (1977).

\bibitem{MIT}
 A. Chodos, R.L. Jaffe, K. Johnson, C.B. Thorn and V.F. Weisskopf,
Phys. Rev. D {\bf 9}, 3471 (1974).

\bibitem{NP}
 G. Chanfray, O. Nachtmann and H.J. Pirner, Phys. Lett. {\bf 147B},
249 (1984); H.B. Nielsen and A. Patkos, Nucl. Phys. {\bf B195}, 137
(1982).

\bibitem{H85}
 R. Goldflam and L. Wilets, Phys. Rev. D {\bf 25}, 1951 (1982).

\bibitem{recoil}
 J.-L. Dethier, R. Goldflam, E.M. Henley and L. Wilets, Phys. Rev. D
{\bf 27}, 2191 (1983); M. Betz and R. Goldflam, Phys. Rev. D {\bf
28}, 2848 (1983).

\bibitem{boost}
 E.G. L\"ubeck, E.M. Henley and L. Wilets, Phys. Rev. D {\bf 35},
2809 (1987).

\bibitem{B851}
 M. Bickeboeller, M.C. Birse, H. Marschall and L. Wilets, Phys.  Rev.
D {\bf 31}, 2892 (1985).

\bibitem{C79}
 A. Casher, H. Neuberger and S. Nussinov, Phys. Rev. D {\bf 20}, 179
(1979).

\bibitem{H81}
 M. Harvey, Nucl. Phys. {\bf A352}, 301 (1981); {\em ibid.} 326.

\bibitem{Ring}
 P. Ring and P. Schuck, {\em The nuclear manybody problem},
(Springer, New York, 1980).

\bibitem{B85}
 M. Bickeboeller, R. Goldflam and L. Wilets, J. Math. Phys.  {\bf
26}, 1810 (1985).

\bibitem{Tang2}
 P. Tang and L. Wilets, J. Math. Phys. {\bf 31}, 1661 (1991).

\bibitem{W84}
 J. Wroldsen and F. Myhrer, Z. Phys. {\bf C25}, 59 (1984).

\bibitem{H16}
 J.F. Donoghue and E. Golowich, Phys. Rev. D {\bf 15}, 3421 (1977).

\bibitem{Pion}
 W. Koepf and L. Wilets, work in progress.

\bibitem{CT75}
 A. Chodos and C.B. Thorn, Phys. Rev. D {\bf 12}, 2733 (1975).

\bibitem{CBM}
  S. Th\'eberge, A.W. Thomas and G.A. Miller, Phys. Rev. D {\bf 22},
2838 (1980); A.W. Thomas, S. Th\'eberge and G.A. Miller, Phys. Rev. D
{\bf 24}, 216 (1981); S. Th\'eberge and A.W. Thomas, Nucl. Phys. {\bf
A393}, 252 (1983).

\bibitem{Corr}
 Fl. Stancu and L. Wilets, {\em Symmetries of Six-quark States
related to the Nucleon-Nucleon Problem}, Proc. Many Body Conference,
Coimbra, Portugal (1993).

\bibitem{BCKS}
 T. Barnes, S. Capstick, M.D. Kovarik and E.S. Swanson, Phys. Rev. C
{\bf 48}, 539 (1993).

\bibitem{Book}
 L. Wilets, {\em Nontopological Solitons}, (World Scientific,
Singapore, 1989).
\end{references}
\end{document}